\documentclass[twocolumn]{aastex61}
\usepackage{epsfig}

\begin{document}
\title{On The Origin of Supernova-Less Long Gamma Ray Bursts}

\author{Shlomo Dado}
\affiliation{Physics Department, Technion, Haifa 32000, Israel}
\author{Arnon Dar}
\affiliation{Physics Department, Technion, Haifa 32000, Israel}

\begin{abstract} 
The fraction of long duration gamma ray bursts (GRBs) without an 
associated bright supernovae (SNe) at small redshifts $(z<0.15)$ 
is comparable to that of GRBs associated with SNe. We show, that their 
X-ray afterglow and those of most of the nearby $(z<1)$ 
GRBs without a confirmed association with SNe, are well reproduced by 
the launch of highly relativistic jets in the SN-less birth of millisecond 
pulsars (MSPs) in neutron star mergers or through phase transition of 
neutron stars to quark stars following mass accretion in compact 
binaries. Such a large fraction of GRBs with pulsar-like afterglow that 
extends to very large redshifts $(z>4)$ favors phase transition of 
neutron stars to quark stars in high mass X-ray binaries (HMXBs) 
rather than neutron stars mergers as their origin.
\end{abstract}

\keywords{gamma-ray burst, jets,  neutron star}

\section{Introduction} 
Gamma ray bursts seem to be divided into two distinct classes, long 
duration soft gamma ray bursts (GRBs) that usually last more than 2 
seconds and short hard bursts (SHBs) that usually last less than 2 
seconds (Norris et al. 1984; Kouveliotou et al. 1993). While there is 
clear photometric (e.g., Dado et al. 2002, Zeh et al. 2004) and 
spectroscopic evidence (e.g., Della Valle et al. 2016 and references 
therein) that a large fraction of the long duration GRBs are produced 
in very bright stripped envelope supernova (SN) explosions of type Ic 
(Galama et al.~1998), there is also evidence that supernova explosions 
are not the only source of long duration GRBs (e.g., Della Valle et 
al. 2006; Fynbo et al. 2006; Gal-Yam et al. 2006). Moreover, in the 
local universe the rate of SN-less GRBs is comparable to that of 
SN-GRBs, as can be seen from Table 1 where all the GRBs with known 
redshift $z<0.15$ are listed. If that is valid universally, then the 
fraction of SN-less GRBs is comparable to that of SN-GRBs. The origin 
of such a large population of SN-less GRBs, which seems to extend to 
very large redshifts is still unknown.

Thirty years ago, Goodman, Dar and Nussinov (1987) suggested that GRBs 
are not Galactic in origin, as was widely believed, but may be 
produced at cosmological distances by highly relativistic 
$e^+e^-\gamma$ fireballs (Goodman 1986) formed by 
neutrino-antineutrino annihilation around merging neutron stars in 
close binaries due to gravitational wave emission and/or around 
compact stars undergoing gravitational collapse due to mass accretion. 
But, shortly after the launch of the Compton gamma ray burst 
observatory (CGRO), it became clear that such neutrino-annihilation 
fireballs are not powerful enough to produce observable GRBs at the 
very large cosmological distances, which were indicated by the CGRO 
observations (Meegan et al. 1992). Consequently, Meszaros and Rees 
(1992) suggested that the $e^+e^-\gamma$ fireballs produced in the 
merger of neutron stars and of neutron stars and stellar black holes 
may be collimated into $e^+e^-\gamma$ jets by funneling through 
surrounding matter. Instead, Shaviv and Dar (1995) suggested that GRBs 
are produced at large cosmological distances by inverse Compton 
scattering (ICS) of ambient light by narrowly collimated jets of 
plasmoids (cannonballs) of ordinary matter rather than by collimated 
fireballs, launched in neutron star mergers, in phase transition of 
neutron stars to quark stars in compact binaries due to mass 
accretion, and in stripped envelope SN explosions. To date, there is 
direct photometric and spectroscopic evidence for production of GRBs 
in stripped envelope SNe of type Ic (see, e.g., Della Valle 2016 and 
references therein), but only indications that SHBs and perhaps 
SN-less GRBs may be produced in neutron star mergers (Tanvir et al. 
2013; Berger et al. 2013;  Hotokezaka et al. 2013; Fong and Berger 
2014, Berger 2014, and Jin et al. 2015, respectively).

There is, however, compelling evidence that both GRBs and SHBs are 
produced by inverse Compton scattering (ICS) of light by highly 
relativistic jets (Shaviv \& Dar 1995).  It includes a large linear 
polarization of the prompt emission, pulse shape and spectral 
evolution during its fast decline phase, and various correlations among 
burst properties (see, e.g., Dar \& De R\'ujula 2000, 2004; Dado et al. 2009a,b; 
Dado \& Dar 2017 and references therein). The main suggested sources of 
such highly relativistic jets, other than SN explosions, include merger of 
neutron stars (Goodman, Dar \& Nussinov 1987), neutron star - black hole 
mergers (Meszaros \& Rees 1992), phase transition of neutron stars to more 
compact objects (quark stars or black holes) following mass accretion 
from a companion star (Dar et al. 1992), and "failed supernovae" (FSN), 
i.e., direct collapse of massive stars to black holes without a supernova 
(e.g., Woosley 1993a,b). These different types of events are likely to 
produce different afterglows and different gravitational wave signals, 
which may be used to identify the origin of SN-less GRBs.

An FSN origin of SN-less GRBs seems to be ruled out: Both FSN-GRBs and 
SN-GRBs are produced by highly relativistic jets emitted in core 
collapse of massive stars which takes place mostly within dense 
molecular clouds. In such environments, their late-time afterglows are 
dominated by synchrotron radiation emitted by the decelerating jets in 
their high density circumburst environment. Such synchrotron afterglows 
satisfy simple closure relations between their late-time spectral and 
temporal behaviors (e.g., Dar \& De R\'ujula 2004; Dado \& Dar 
2013,2017) independent of the jet origin. However, the observed late-time 
afterglows of SN-less GRBs that are expected to be powered by millisecond 
pulsars (MSPs) are different from those of 
SN-GRBs (e.g., Cano et al. 2017) and do not satisfy the standard closure 
relations of SN-GRBs afterglows.

Among the other alternative sources of SN-less GRBs, either merger or 
phase transition of neutron stars within globular clusters (GCs) of very high 
density interstellar light and very low density interstellar medium 
(ISM) may explain the observed late-time X-ray afterglow of SN-less 
GRBs. Indeed, in this paper we show that the light curves of the X-ray 
afterglow of the well sampled nearby SN-less GRBs measured with the 
Swift X-ray telescope (Evans et al. 2007, 2009) are well explained by 
ICS of GC light by relativistic jets launched in the birth of an MSP, 
which is taken over by an X-ray afterglow powered by the spin down of 
the newly born MSP (Dai and Lu 1998a,b; Zhang and Meszaros 2001; Dai et 
al. 2006; Metzger et al. 2008). Late time afterglows powered by MSPs 
rule out neutron star - black hole mergers or phase transition of 
neutron stars to black holes as the origin of SN-less GRBs. However, 
they do not distinguish between neutron stars merger origin and 
collapse of neutron stars to quark stars following mass accretion. But, 
in this paper, we also show that the fraction of GRBs with a very large 
measured redshift ($z>4$), which have "MSP-like" late-time X-ray 
afterglow (12 0ut of 24), is like that observed at very low redshifts 
($z<0.15$). It implies the same redshift dependence of the production 
rate of SN-less GRBs and SN-GRBs. It favors mass accretion on neutron 
stars in high mass (short lived) X-ray binaries (HMXBs) as the origin 
of SN-less GRBs (Dar 1998) rather than merger of neutron stars in compact 
binaries by gravitational wave emission which takes time comparable to 
the age of the universe.

\section{The fast decline phase of prompt emission}

Consider first a jet which is a succession of highly relativistic 
plasmoids (cannonballs) launched during the birth of MSP in neutron 
star merger or phase transition in a close binary within a collapsed 
core of a globular cluster. Let $\gamma$ be the 
bulk motion Lorentz factor of such a cannonball (CB) and 
$\delta= 1/(1-\beta\,cos\theta)$ be its Doppler factor when viewed at 
an angle $\theta$ relative to its direction of motion. As long as the 
CB moves inside the collapsed core, ICS of its 
nearly isotropic light by the plasmoid electrons produces a 
light curve proportional to the light density along its trajectory. As 
the highly relativistic CB leaves the collapsed  core and moves away, 
the angle between the CB motion and the intercepted GC photons decreases 
with distance $r$ from the GC core  like $r/R_c$, where $R_c$ is the 
radius of the collapsed core of the GC.
The transverse momentum of the intercepted GC photons in the 
CB rest frame is unchanged, while their parallel component is strongly 
red shifted. Hence, in the plasmoid rest frame, the energy $\epsilon$ 
of an intercepted GC photon is Lorentz transformed to an energy 
$(R_c/r)\,\epsilon$. This energy is Doppler boosted by ICS to an energy 
$E\approx R\,\epsilon /(1+z)\, r$ when observed at a viewing angle $\theta$ in 
the rest frame of a distant observer. Time aberration yields a distance 
$r=c\,\gamma\,\delta\,t/(1+z)$ of the CB from the GC at an observer time 
$t$. Hence, for a GC light with a bremsstrahlung-like spectrum, 
$\epsilon^2\,dn/d\epsilon\propto exp(-\epsilon/\epsilon_p))$ and a 
density $n\propto 1/r^2$ at $r >> R_c$, the decay of the energy density 
flux $F(t)$ from ICS of a GC light by a highly relativistic plasmoid launched 
at $t=0$ is given by (Dado and Dar 2017),
\begin{equation} 
F(t)=E^2\, dN/dE \propto exp(-t/\tau) 
\end{equation}
with $\tau(E)\approx 100(R_c/pc)(\epsilon_p/eV)/c(E/keV)(\gamma /10^3)$
s. This decay rate of the prompt emission depends strongly on 
the observed energy band. It is valid only when $r>>R_c$, i.e., for 
$t>>t_{es}$, where $t_{es}$ is the escape time of the highly 
relativistic plasmoids from the GC core at redshift $z$, as observed by 
a distant observer from a viewing angle $\theta \approx 1/\gamma$. 
The $E$-dependence of $\tau$ produces the observed (Evans et al.
2007, 2009) fast spectral softening as function of time 
during the exponential  decline of the prompt emission for $t>t_{es}$.
Due to time aberration, $t_{es}\approx (1 + z)\,R_c/\gamma\, \delta\, c$. 
The median value of the obsered radii of collapsed cores of GCs is 
$R_c\sim 1$ pc. Hence, $t_{es}\approx (1 + z)\,50$ s, for
a plasmoid with $\delta\approx \gamma\approx 500$, which is similar to 
the typical time that the Swift-XRT usually begins its follow-up 
observations. 

The successive emission of plasmoids during a GRB and/or a bumpy light 
density in the globular cluster yields a bumpy light curve, with resolved
peaks  at early times because of large photon statistics. 
Later, when the luminosity decreases, the larger time bins (and perhaps 
merger of CBs) yield smoothly appearing light curves. The 
smoothed X-ray light-curve during the fast decay phase of the prompt 
ICS emission from the jet is expected to be described roughly by
\begin{equation}
F_j(t)\approx {a\, F_j(0) \over a + (t/\tau) exp(t/\tau)}
\end{equation}
where $a$ is an adjustable parameter.
The radius of large GCs can reach  $R\approx 25$ pc.
If the GRB occurred within the collapsed core (cc) of a large GC,
ICS of the quasi isotropic GC light by the CB 
inside the GC can take over during the fast decay of the contribution from the 
collapsed core, and produce a second bump with a shape given also 
by Eq.(2), but with a much smaller $F_j(0)$, 
$\tau_{GC}\gg\tau_{cc}$, and $t_{es}\approx (1+z)\times 10^4 (R/25\ pc)(\gamma/500)^2$, 
which is taken over during its fast decay phase 
by the MSP contribution.

The light-curve of a late GRB pulse, or a flare during the decline of the 
prompt emission, which is produced by ICS of ambient light by a CB 
launched at a time $t_{i,0}$, is given by (e.g., Dar and De R\'jula 2004) 
\begin{equation}
 F_i(t_i)\approx F_i{\Delta}\,{4\,(t_i/\Delta_i)^2 \over
                ((t_i/\Delta_i)^2+1)^2}
\end{equation}
where $t_i = t-t_{i,0}$, is the time after $t_{i,0}$ 
the beginning time  of pulse i whose peak value 
is at $t_i=\Delta_i$.
A smooth interpolation from this early time behavior 
to the late time behavior given by Eq.(1), is provided by
\begin{equation}
F_i(t)= F_{p,i}\,{2.528\,(t_i/\Delta_i)^2\, exp(-t_i/\tau_i)
       \over ((t_i/\Delta_i)^2+1)^2\,(1 - exp(-\Delta_i/t_i))}
\end{equation}
where $2.528=4(1-e^{-1})$.

\section{Afterglows powered by MSPs}

Newly born neutron stars seem to be surrounded by nearby plerions, 
which absorb their emitted radiation, winds, and highly relativistic 
particles, and convert them to luminous energy, mostly in the X-ray 
band.  If the power source is the spin down of a newly born neutron 
star with a period $P$, then, in a steady state  it generates 
a plerion luminosity $L(t)= 2\,\pi^2\,I\,\dot{P}/P^3$. For canonical neutron 
stars of a radius $R=10^6$ cm and a mass $M\approx M_{Ch}$, where 
$M_{Ch}\approx 1.4 M_\odot$ is the Chandrasekhar mass limit of white 
dwarfs, $I=(2/5)M_{Ch} R^2\approx 1.12\times 10^{45} g\, cm^2$. 

Observations of young X-ray pulsars indicate that 
the change of their period  during their spin down 
satisfies to a good approximation,    
\begin{equation}
P\dot{P}=K
\end{equation}
where $K$ is time independent constant (with time-dimension). 
Such a behavior is expected from pulsars which spin down by 
magnetic dipole radiation (MDP) from a magnetic dipole aligned at an 
angle $\alpha>0$ relative to the rotation axis, and stays constant in 
time (in the pulsar rest frame) during spin down (e.g., Manchester
\& Taylor 1977). Observations, however, indicate that 
$P\,\dot{P}$ remains constant to a good approximation also when 
other spin down mechanisms, such as emission of winds and/or highly 
relativistic particles, contribute, or even dominate the spin-down
of pulsars. This follows from the fact that the  age estimate 
$t=[P^2(t)-P^2(0)]/2\,K$  and the braking relation 
$d^2P/dt^2\approx -K^2/P^3$ (Manchester \& Taylor, 1977)
seem to be satisfied also by young X-ray pulsars:
The age relation has been tested by comparing the age obtained from 
measurements of $P$ and $\dot{P}$ of young pulsars  to the known 
ages of their parent supernovae  (historical supernovae) or to their ages 
obtained from measurements of the distance and proper motion of the
young pulsars relative to the centers of the supernova remnants where 
they were born. The braking relation, which is very difficult to 
test over a human time scale, has been verified only in few young 
X-ray pulsars where it yielded a braking index near the expected 
value 3 (see, e.g., Archibald et al. 2016 and references therein).

We shall assume that Eq.(5) is valid approximately for the power 
supply by the newly born millisecond neutron stars in SN-less GRBs
during the first few days after their birth. 
It then follows from Eq. (5) that,
\begin{equation}
L_{ps}(t)=L_{ps}(0)\,(1 + t/t_b)^{-2}
\end{equation}
with $t_b=P_0/2\dot{P}_0$ and $P_0=P(t=0)$. The afterglow of a  
GRB at redshift $z$  that is powered by such a luminosity has a local
energy flux density $F(t)=L(t)/4\,\pi\ D_L^2$, where $D_L$ is the 
luminosity distance of the GRB at redshift $z$, i.e.,
\begin{equation} 
F_{ps}(t)=F_{ps}(0)\,(1 + t/t_b)^{-2}\, .
\end{equation}
If only a fraction $\eta$ of the rotational energy is converted to 
X-rays, then  
\begin{equation}
P_0={1\over D_L}\sqrt{{(1+z)\,\pi\, \eta_x\, I \over  2\,F_{ps,x}(0)\,t_b}},
\end{equation}
and its time derivative is  $\dot{P}_0=P_0/2\,t_b$. 

Under the assumptions that the spin down of the newly born 
neutron star is dominated by radiation from a magnetic dipole 
${\bf m}$, which is constant in time and is aligned at a fixed 
angle $\alpha$ relative to its rotation axis, the peak surface 
value of the dipole magnetic field $B_p=2\,m/R^3$ is at the 
magnetic poles and it satisfies (Manchester \& Taylor 1977)
\begin{equation} 
B_p\,sin\alpha \approx 6.8\times 10^{19}[P\dot{P}/s]^{1/2}\, gauss. 
\end{equation} 
However, the wide use of Eq.(9) for  estimatibg $B_p$, cannot be 
trusted  because it assumes that there is no magnetic field 
decay, that the spin down is in a perfect vacuum, that there 
are no other spin down mechanisms in operation, such as emission 
of a wind, relativistic particles, and gravitational waves, 
and that other sources such as thermal
energy, stress energy, phase transition and gravitational 
contraction can be ignored.

\section{Comparison with observations} 
Eq. (2) and (7) can be combined to yield the canonical light curve of 
the decay of the prompt emission phase of SN-less GRBs taken over by 
a MSP powered afterglow,
\begin{equation} 
F(t)\approx {a\, F(0) \over a + (t/\tau) exp(t/\tau)} + 
                {F_{ps}\over (1+t/t_b)^2}\,.
\end{equation}
Figures 1-4, show the X-ray light curves of the four nearest SN-less GRBs, 
051109B, 080517, 060614, and 050826, with known redshifts and a well-sampled 
X-ray afterglows, which were measured with the Swift-XRT and are reported in 
the Swift-XRT GRB light curve repository (Evans et al. 2007,2009). Also shown 
are their best fit light curves as give by Eq.(10) with the minimal number of 
parameters. The best fit values of these parameters (2 for the late-time MSP 
contribution and 2 or 3 for the jet contribution) are listed in Table 2. 

We have also verified that all other well sampled X-ray afterglows of GRBs 
with redshift $z<1$, which were reported in the Swift-XRT and have no 
evidence for an associated SN, despite their relative proximity, could be 
well fit by Eq.(10) (18 additional GRBs in Figures 5-7), or 
by Eq.(4) plus Eq.(7)  (6 additional GRBs in Figure 8) 
with flaring activity during the decay phase of their prompt emission.

Moreover, Figures 9 and 10 show the well-sampled 
X-ray afterglows of 12 GRBs out of 24 with a known large redshift ($z>4$) that 
were reported in the Swift-XRT GRB light curve repository (Evans et al. 
2007,2009) and could be well fit by a jet + MSP light curve
with the parameters listed in Table 4 as shown in these figures.

\section{Discussion}

Fast rotating milliseconfd neutron stars can be produced in SN explosions, 
in merger of neutron stars and by spin up of neutron stars by mass 
accretion in compact binaries. We have conjectured that the 
spin down of such newly born young millisecond neutron stars  satisfy 
to a good approximation, $P\,\dot{P}=K$ where $K$ is constant in time 
during the first few days after birth.
This conjecture was motivated by the fact that 
the braking relation $d^2P/dt^2\approx -\dot{P}^2/P\approx -K^2/P^3$ 
and the age estimate $t_{age}\approx P/2\,\dot{P}$, which follow from the 
relation $P\,\dot{P}\approx const$ seem to be satisfied to a good 
approximation by young X-ray MSPs: Their braking index derived from 
precise measurements of $P(t)$ of a few young MSPs is near 3, and the 
age relation $t_{age}=(P^2-P_0^2)/2\,P\, \dot{P}$ of young pulsars seem 
to agree with the age of their parent historical supernovae or their 
age extracted from their observed distance and proper motion relative 
to the centers of their parent supernovae.

We have shown that all the well sampled light curves of the  X-ray 
afterglow of the nearby SN-less GRBs measured with the Swift X-ray 
telescope are those expected from the launch of highly  relativistic jets  in 
the birth of MSPs within an environment of a very dense light and a very 
low ISM density. Such an environment exists in stellar-rich GCs within or 
without a collapsed core.  We have also shown that roughly half of all 
GRBs with well sampled X-ray afterglow, where  SN-GRB 
association has not or could not be confirmed, could be well 
explained by ICS of light by relativistic jets launched in the birth of an 
MSP within a GC like environment.

The late time behavior of the light curves of the afterglows of SN-less 
GRBs is very different from that of SN-GRBs. In SN-GRBs, the light curves 
are well explained by synchrotron radiation from the deceleration of 
highly relativistic jets in dense molecular clouds within spiral galaxies, 
where most massive star formation and core collapse supernovae of type Ic 
take place.  The late-time light curves of SN-less GRBs can neither be 
explained well by synchrotron radiation from the deceleration of highly 
relativistic jets nor do they satisfy the closure relations which are 
satisfied by the late time afterglows of SN-GRBs.
 
The MSP birth periods obtained from
our fits to the late time X-ray afterglow of SN-less GRBs with known 
redshift are all  well above the minimal classical period 
$2\pi\,R/c\approx 0.2$ ms of fast rotating canonical neutron stars. 

Late-time afterglows powered by MSPs rule out neutron star - black hole 
mergers or phase transition of neutron stars to black holes as the origin 
of SN-less GRBs. However, they do not distinguish between neutron stars 
merger origin or collapse of neutron stars to quark stars following mass 
accretion.

SHBs are observed near/within both spiral and elliptical galaxies 
(probably within the collapsed core of stellar rich GCs), while long soft 
GRBs seems to take place only in spiral galaxies mainly within star 
formation regions. This suggest that SHBs and SN-less GRBs have different 
origins: We have found that the fraction of GRBs with very large measured 
redshifts ($z>4$), which have "MSP-like" late-time X-ray afterglow is not 
different from that observed at very low redshifts ($z<0.15$). This 
implies a similar redshift dependence of the rate of SN-less GRBs and 
SN-GRBs, is proportional to the global star formation rate 
when beaming and threshold effects are taken into consideration
(Dado \& Dar 2013). Neutron star merger via 
gravitational wave emission in compact binaries, usually takes a very long 
time comparable to the age of the universe. If SN-less GRBs were formed by 
neutron stars merger in compact binaries due to gravitational wave 
emission, it would have implied a a much lower fraction of SN-less GRBs at 
very large redshifts, i.e., a much smaller fraction of GRBs which have a 
late-time X-ray afterglow that appears to be powered by MSP. That is not 
observed.  We have found that out of the 30 GRBs with a measured redshift 
$z>4$  about half have a late time X-ray afterglows similar to those of 
the SN-less GRBs at low-z ($z<0.15$), which consist about half of the low 
$z$ GRBs. This favors phase transition of neutron stars to quark stars due 
to mass accretion in HMXBs as the origin of SN-less GRBs over merger of 
neutron stars due to gravitational wave emission in compact binaries. Such 
HMXBs may be present mostly in GCs and star formtion regions in spiral galaxies 
where winds and SN ejecta have already swept away the initial high density 
ISM.
 
\begin{table*}
\centering
\caption{~~~~~The SN-GRBs and SN-less GRBs with known redshift $z\leq 
0.15$\,.}

\begin{tabular}{l l l l l l}
\hline
\hline
~~~SN~~&SN-GRB~&~~z~&SN-less GRB&~~~~~Ref.~~\\
\hline
1998bw & 980425~& 0.0085 &         & Galama et al. 1998~~~~ \\
       &        & 0.013~ & 111005A & Michalowski et al. 2016\\
2006ag & 060218~& 0.0331 &         & Pian et al. 2006~~~~~~~\\
2010bh & 100316D& 0.059~ &         & Starling et al. 2011~~~\\
       &        & 0.080 ?& 051109B & Perley et al. 2006~~~~~~\\
       &        & 0.0809~& 060505~ & Fynbo et al. 2006~~~~~~\\
       &        & 0.0890 & 080517~ & Stanway et al. 2015~~~~\\
       &        & 0.125  & 060614~ & Fynbo et al. 2006~~~~~ \\
2013dx & 130702A& 0.145  &         & D'Elia et al. 2015~~~~~\\
2016jca& 161219B& 0.1475 &         & Cano et al. 2017~~~~~~~\\
\hline
\end{tabular}
\end{table*}

\begin{figure}[]
\centering
\epsfig{file=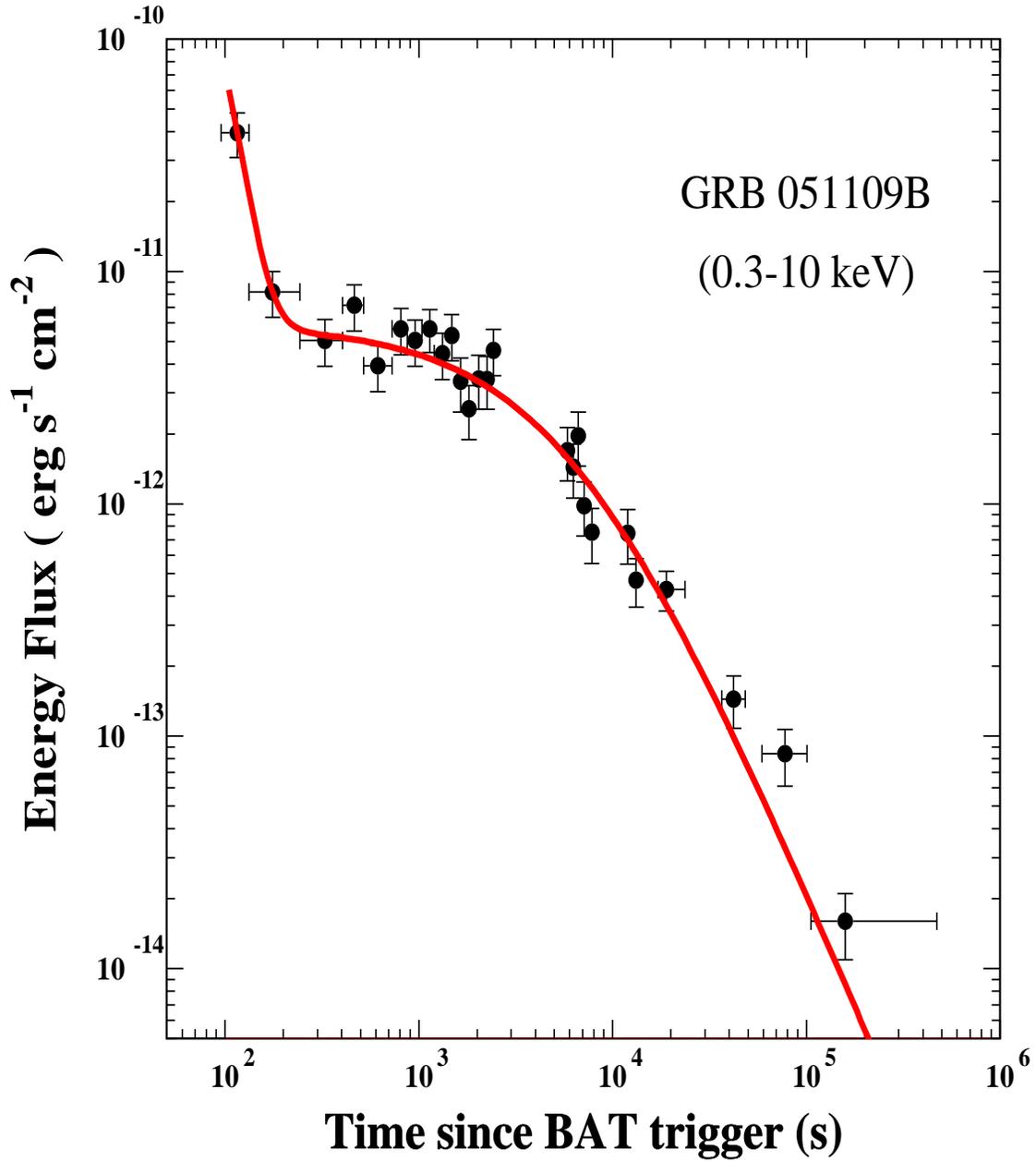,width=16.cm,height=18.cm}
\caption{The X-ray light curve of the SN-less GRB 051109B at the 
(uncertain) redshift $z=0.08$
measured with Swift BAT and XRT telescopes  and
reported by Troja et al. (2006) 
and in the Swift-XRT GRB light curve 
repository (Evans et al. 2007,2009).
The line is the best fit light curve as given by Eq.(10)
with the parameters listed in Table 2.}     
\label{Fig1}                                                                       
\end{figure}

\begin{figure}[]
\centering
\epsfig{file=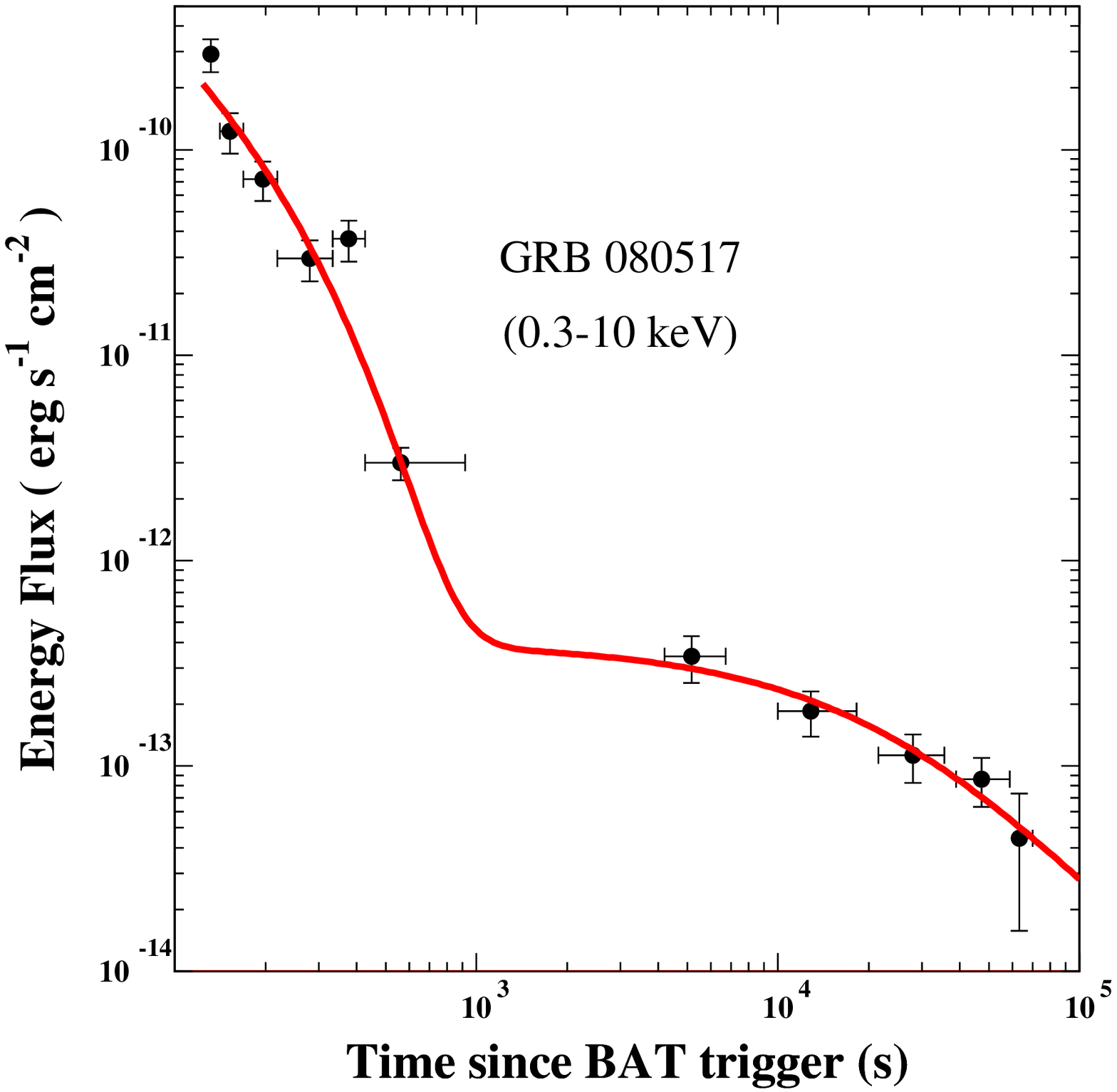,width=16.cm,height=18.cm}
\caption{The X-ray light curve of the SN-less GRB080517 
at redshift $z=0.0809$ reported in the Swift-XRT GRB 
light curve repository (Evans et al. 2007,2009). 
The line is the best fit light curve as given by Eq.(10)
with the parameters listed in Table 2.}     
\label{Fig2}                                                                       
\end{figure}

\begin{figure}[]
\centering
\epsfig{file=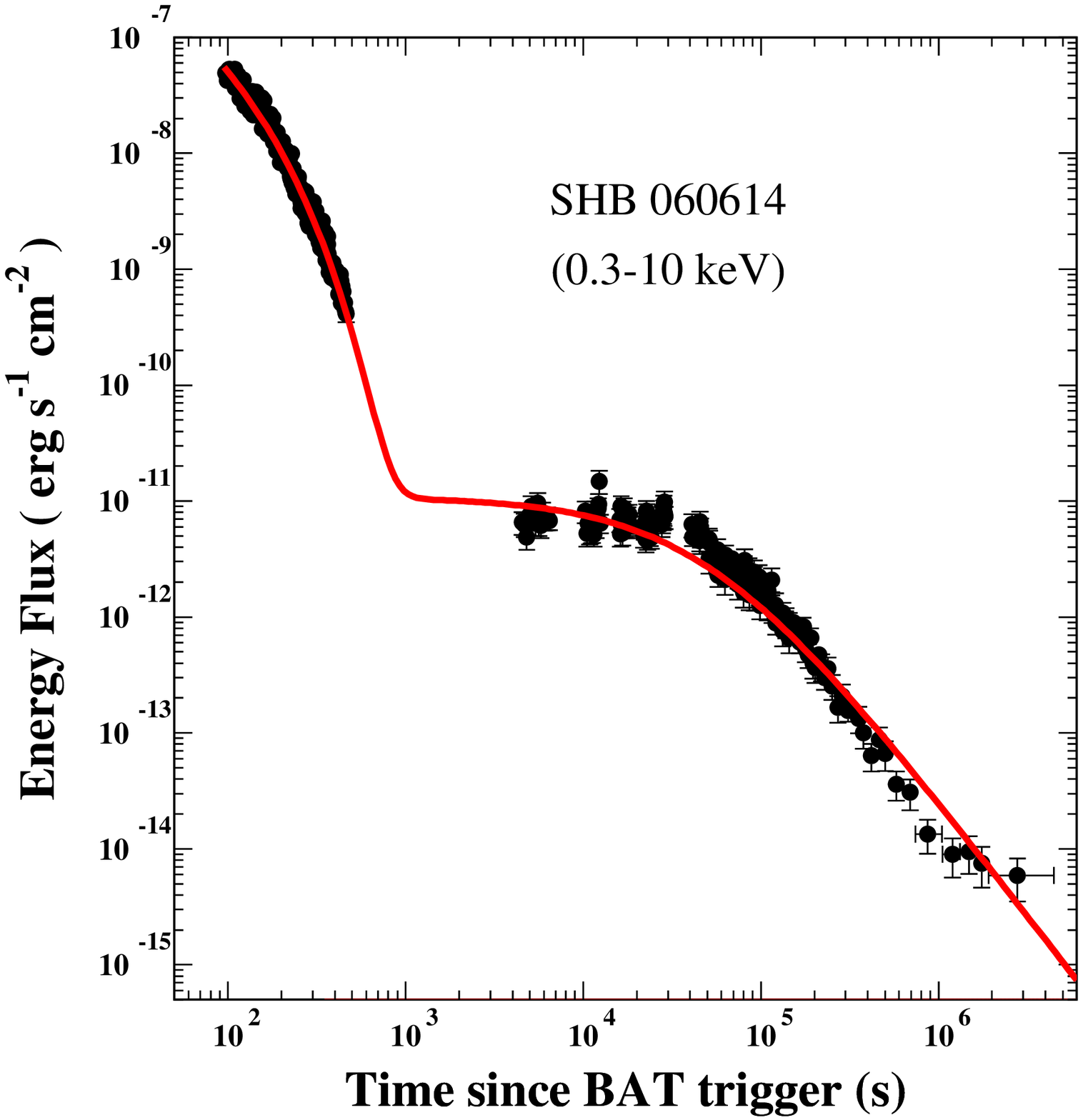,width=16.cm,height=18.cm}    
\caption{The X-ray light curve of the SN-less GRB 060614 
at redshift $z=0.125$
reported in the Swift-XRT GRB light curve repository (Evans et al.
2007,2009). The line is the best fit light curve as given by 
Eq.(10) with the parameters listed in Table 2.} 
\label{Fig3}
\end{figure} 

\begin{figure}[]
\centering
\epsfig{file=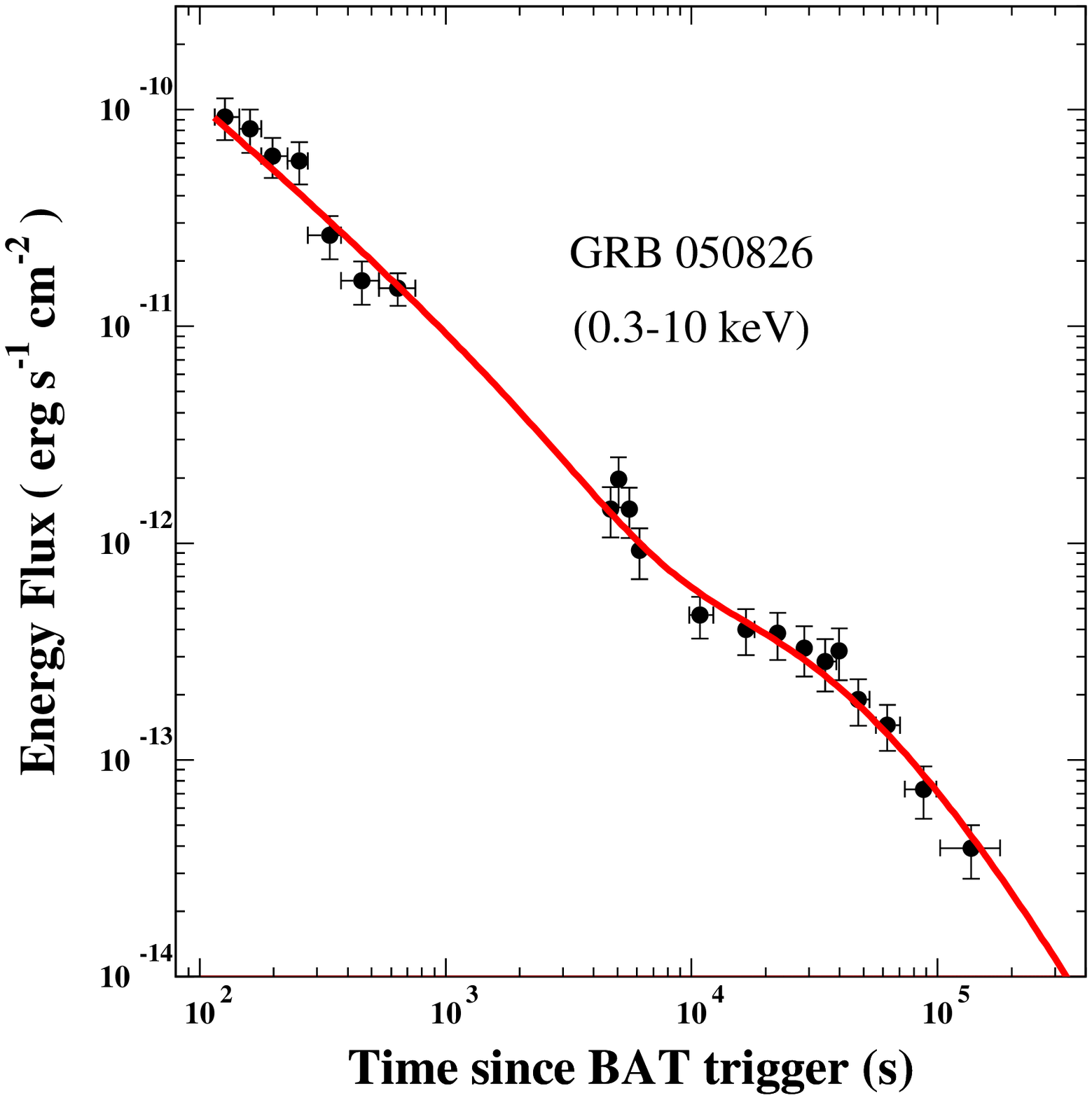,width=16.cm,height=18.cm}
\caption{The X-ray light curve of the SN-less GRB050826 
at resshift $z=0.297$  
reported in the Swift-XRT GRB light curve repository (Evans 
et al. 2007,2009). 
The line  is the best fit light curve as given by Eq.(10)
with the parameters listed in Table 2.}     
\label{Fig4}                                                                       
\end{figure}

\begin{figure}[]
\centering
\vbox{
\hbox{
\epsfig{file=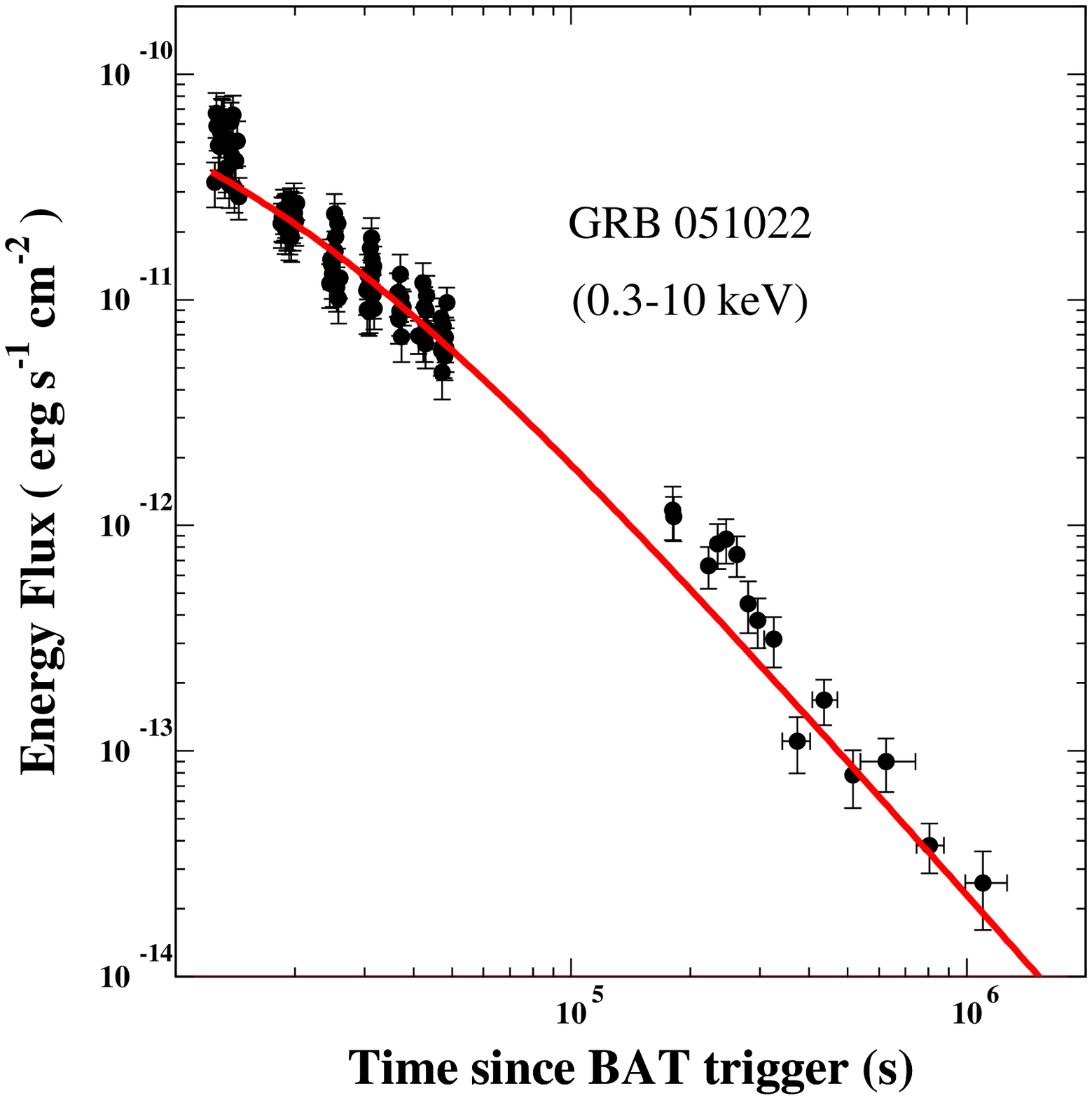,width=8.cm,height=8.cm}
\epsfig{file=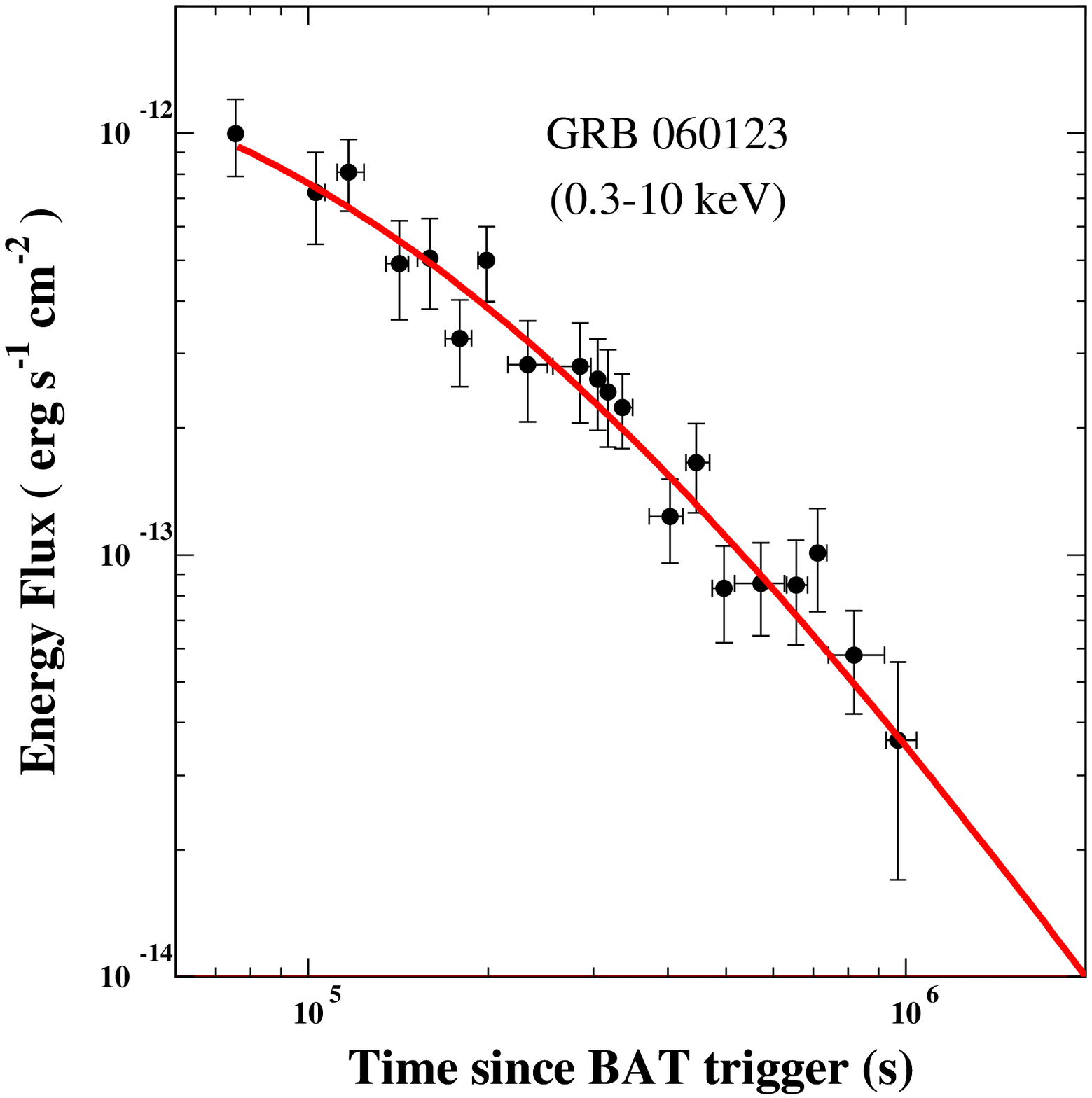,width=8.cm,height=8.cm}
}}
\vbox{
\hbox{ 
\epsfig{file=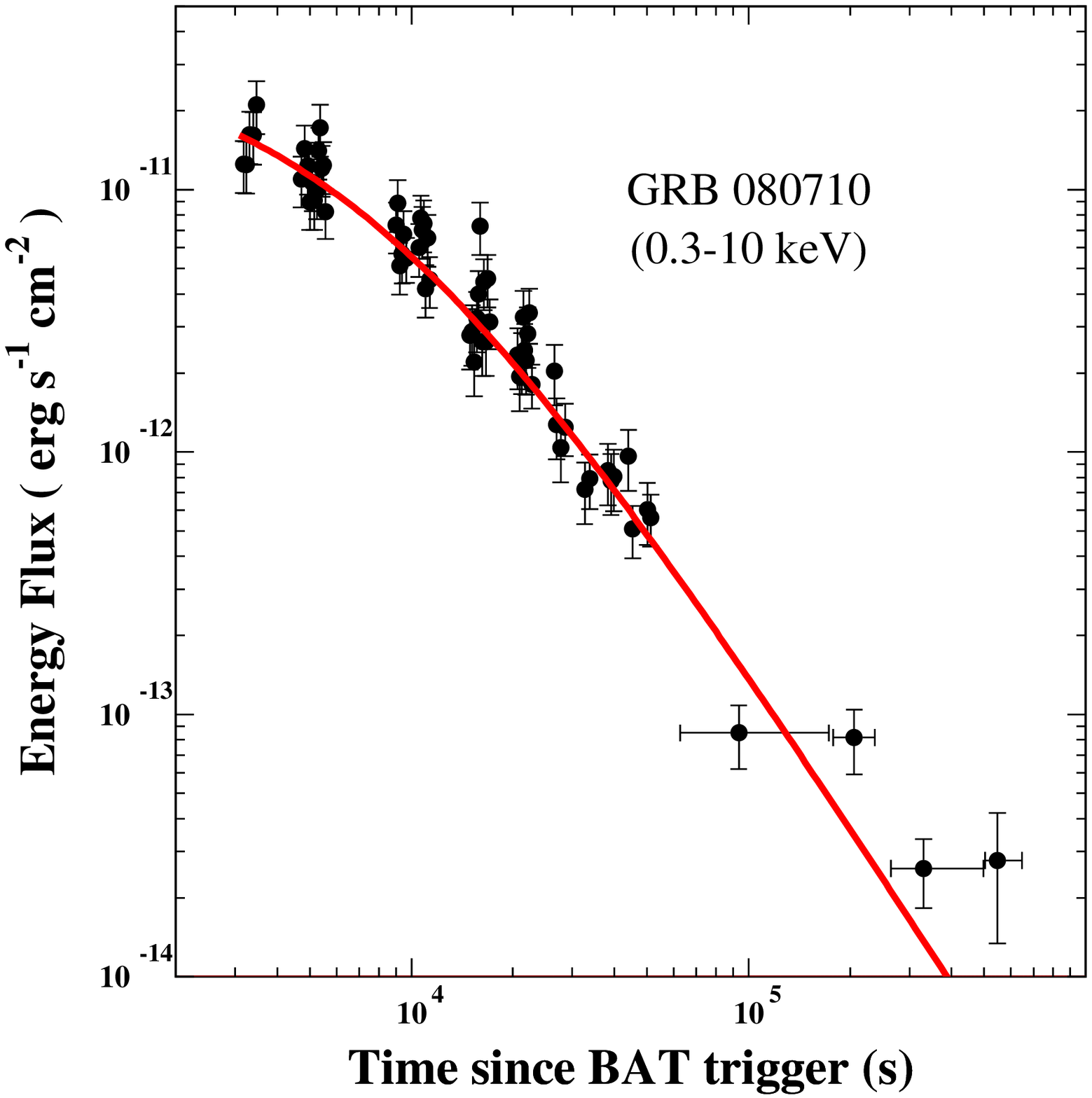,width=8.cm,height=8.cm}
\epsfig{file=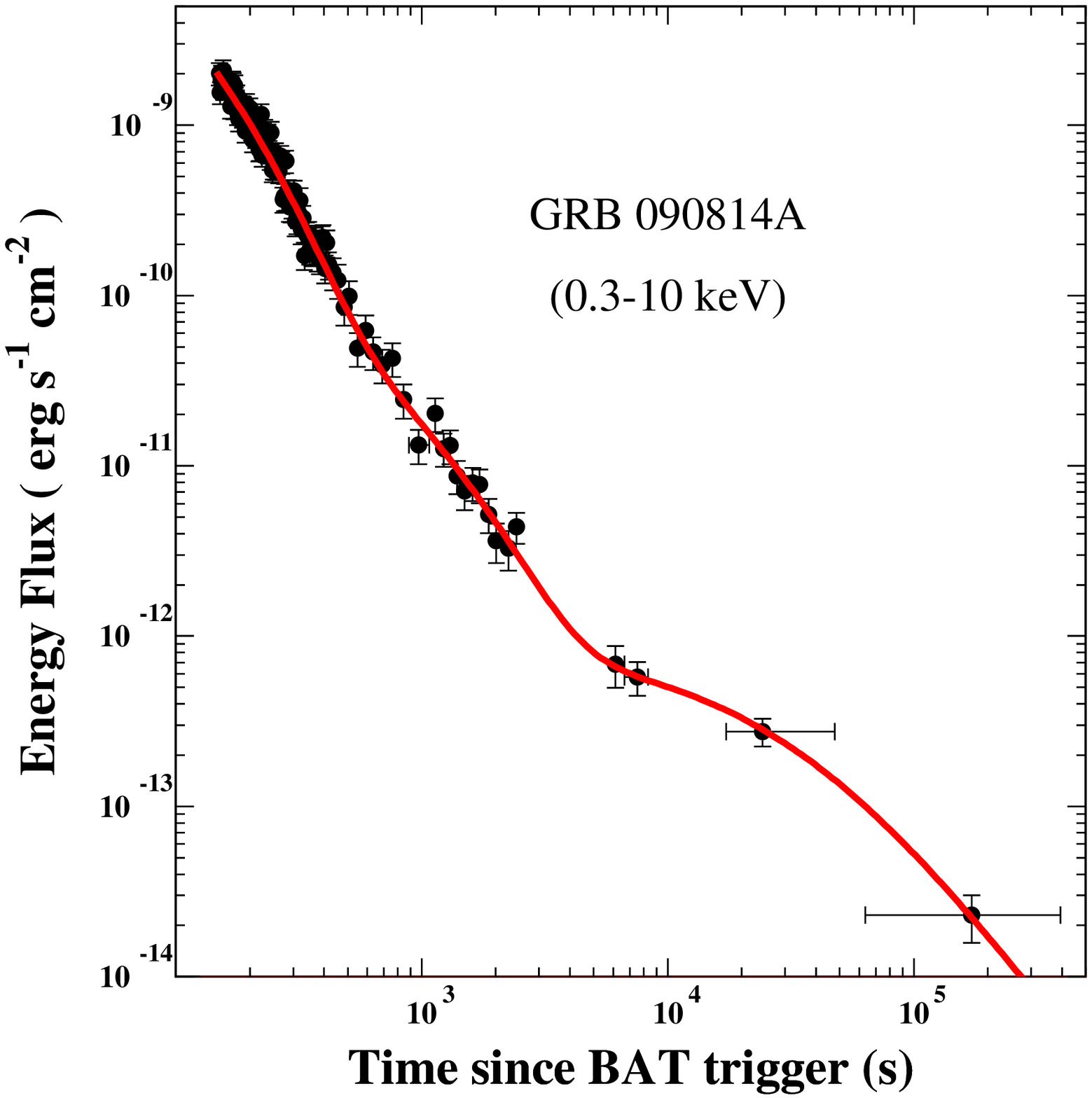,width=8.cm,height=8.cm}
}}
\vbox{
\hbox{
\epsfig{file=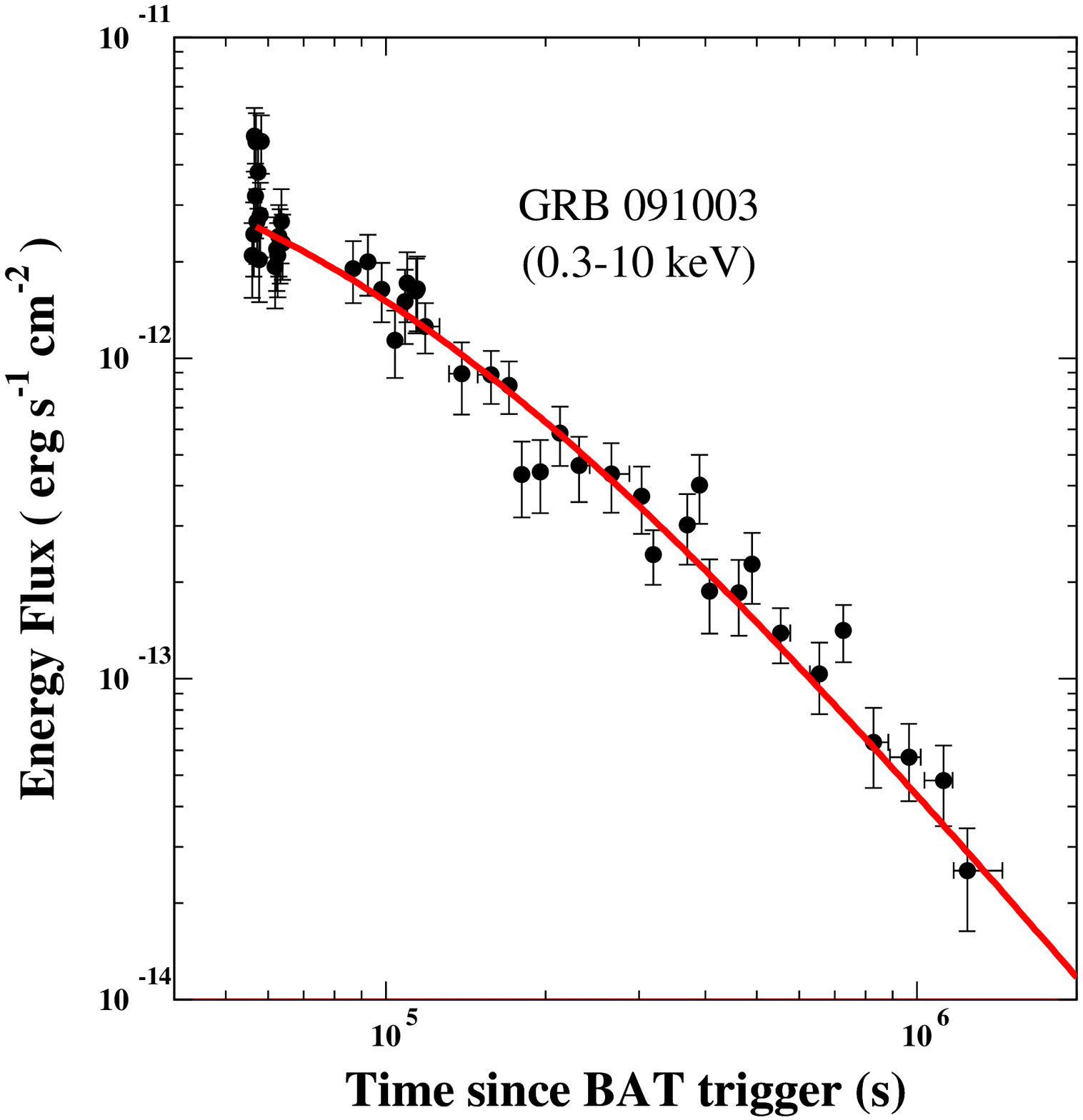,width=8.cm,height=8.cm} 
\epsfig{file=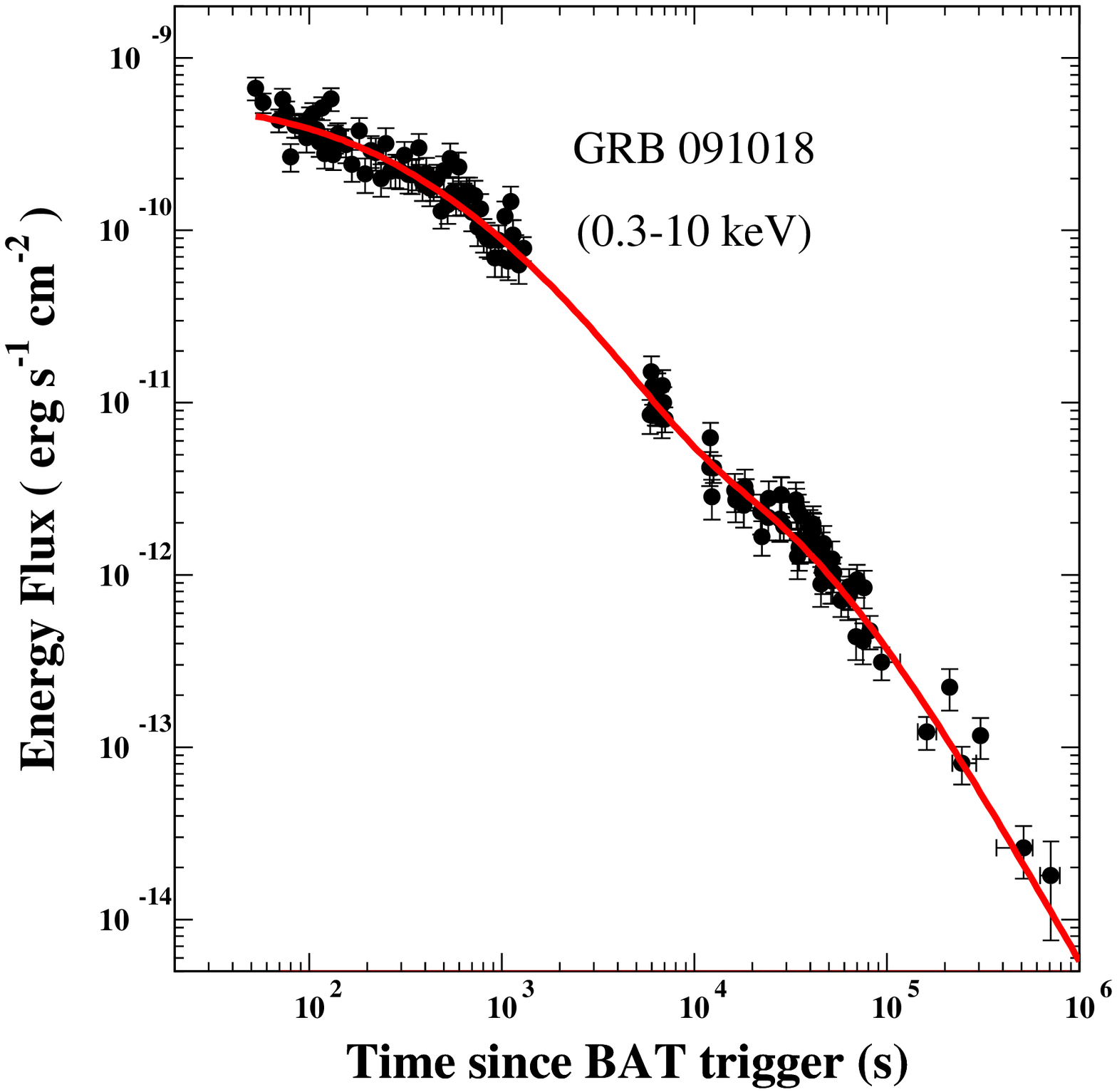,width=8.cm,height=8.cm}
}}
\caption{The X-ray light curve of the nearby ($0.15<z<1) $SN-less GRBs 
051022, 060123, 080710,  090814A, 091003, and 091018  
reported in the Swift-XRT GRB light curve repository 
(Evans et al. 2007,2009)
and their best fit light curve for a jet contribution taken over
by MSP powered emission, as given by Eq.(10) with the parameters
listed in Table 3.}
\label{Fig5}
\end{figure}

\begin{figure}[]
\centering 
\vspace{-1cm} 
\vbox{ 
\hbox{ 
\epsfig{file=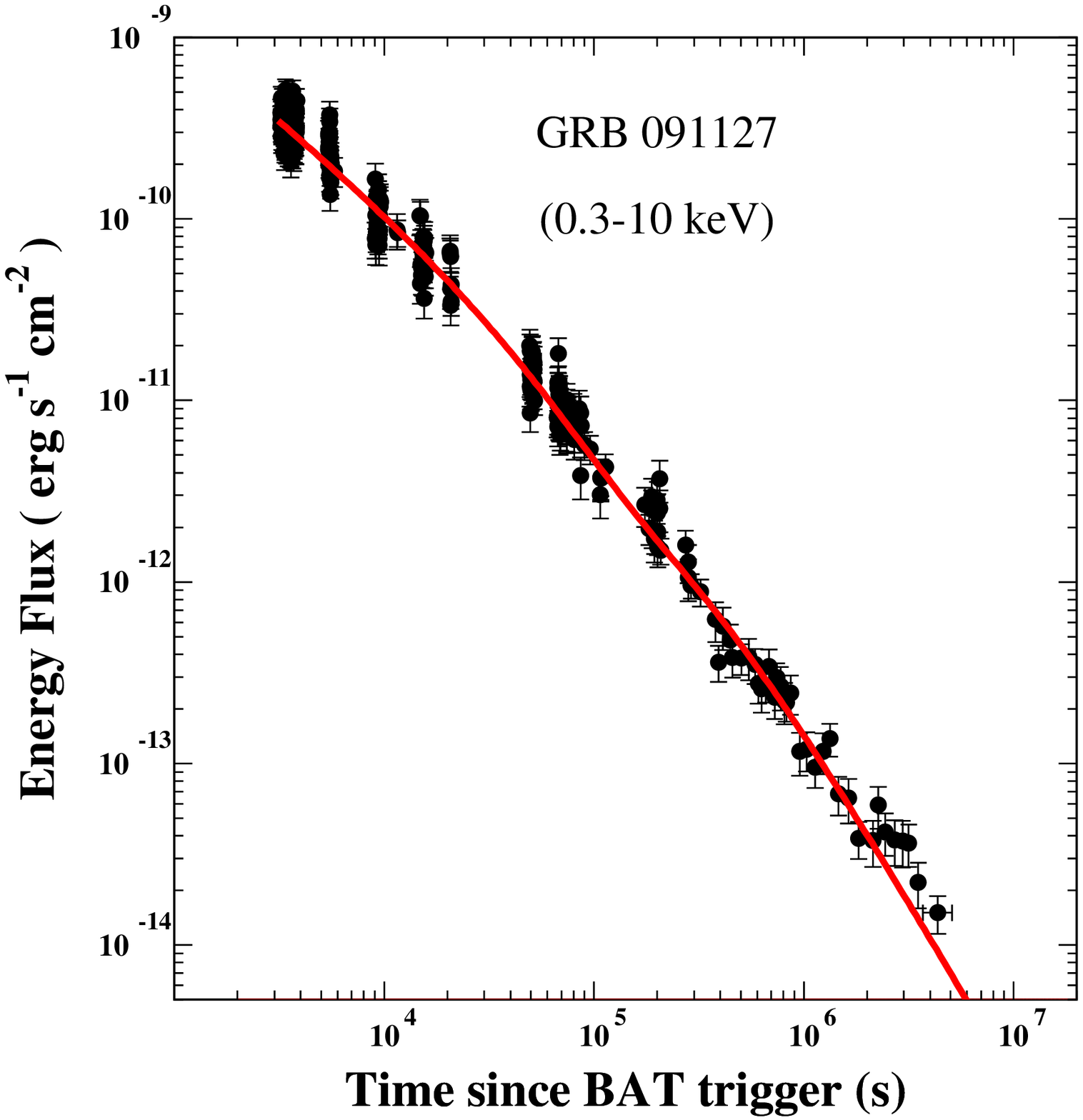,width=8.cm,height=8.cm}
\epsfig{file=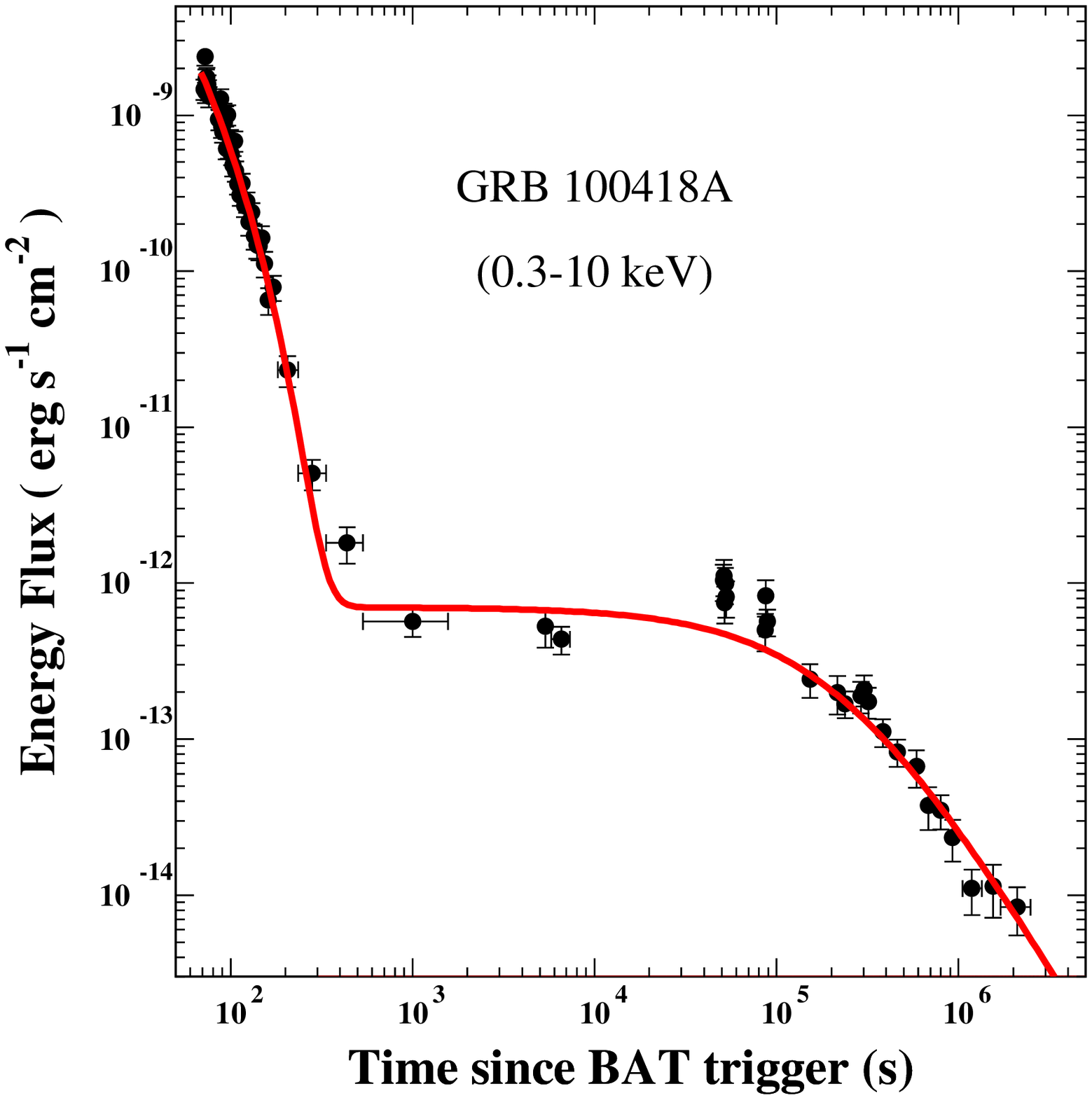,width=8.cm,height=8.cm} 
}} 
\vbox{ 
\hbox{
\epsfig{file=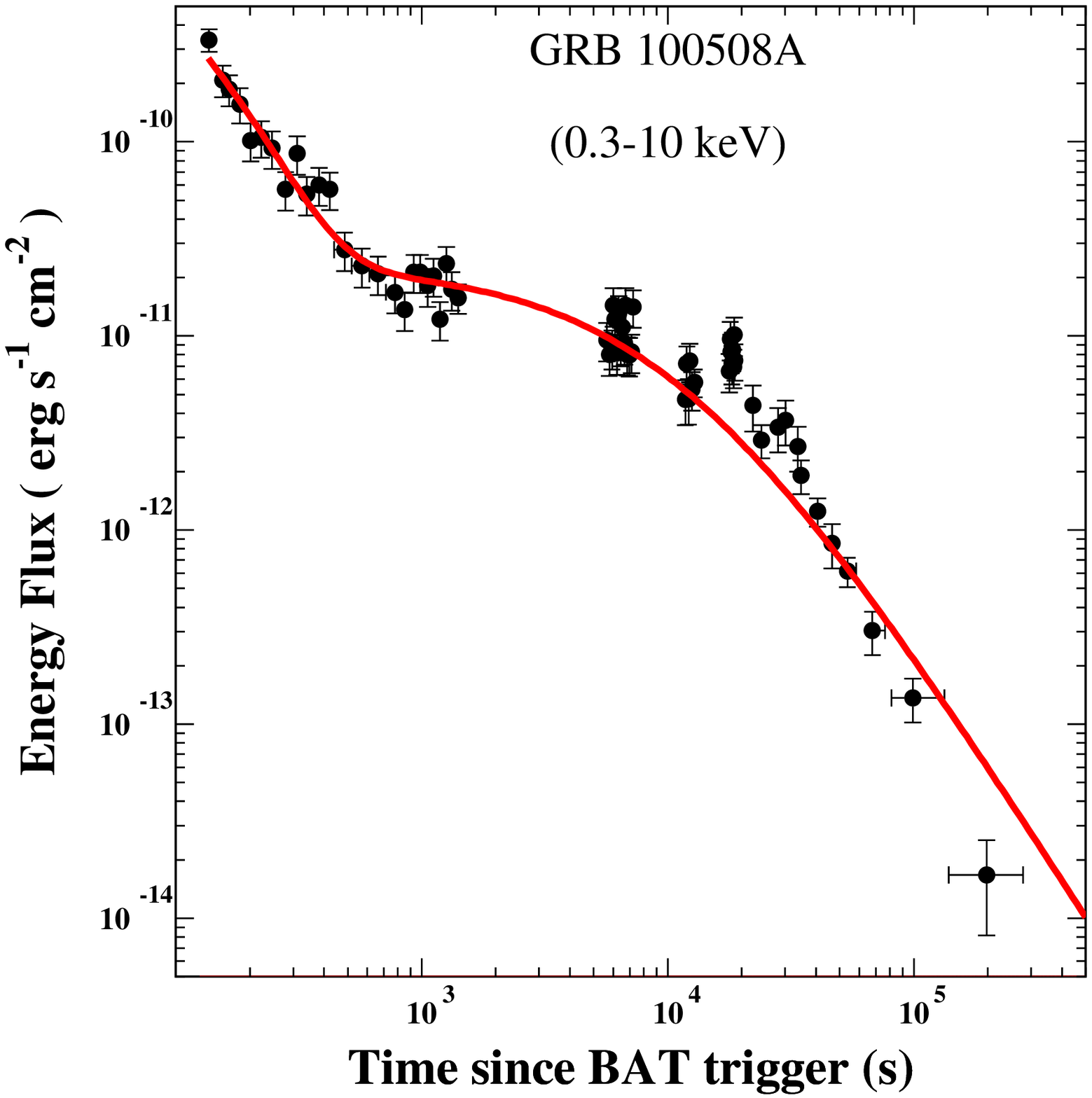,width=8.cm,height=8.cm} 
\epsfig{file=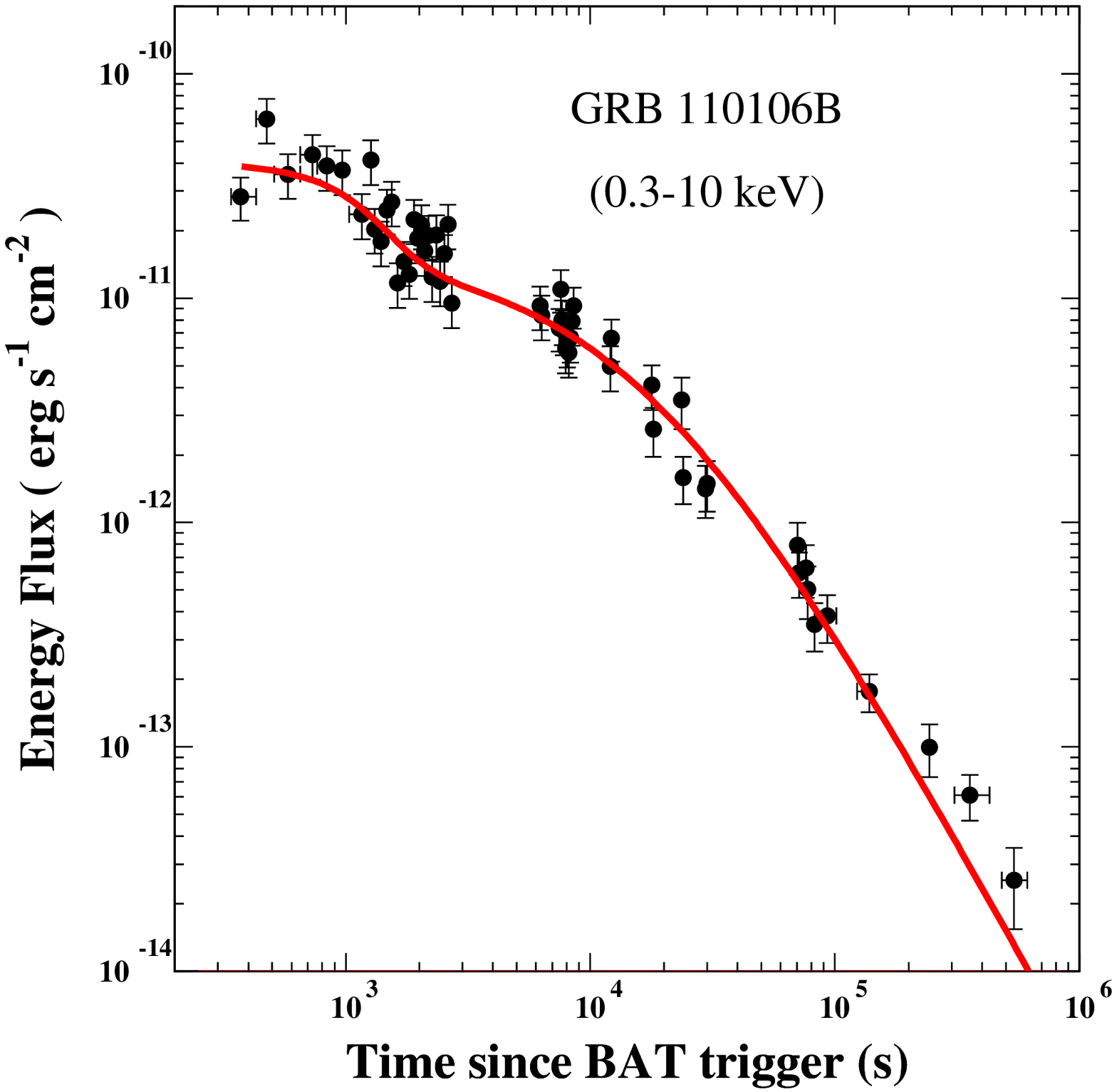,width=8.cm,height=8.cm} 
}} 
\vbox{ 
\hbox{ 
\epsfig{file=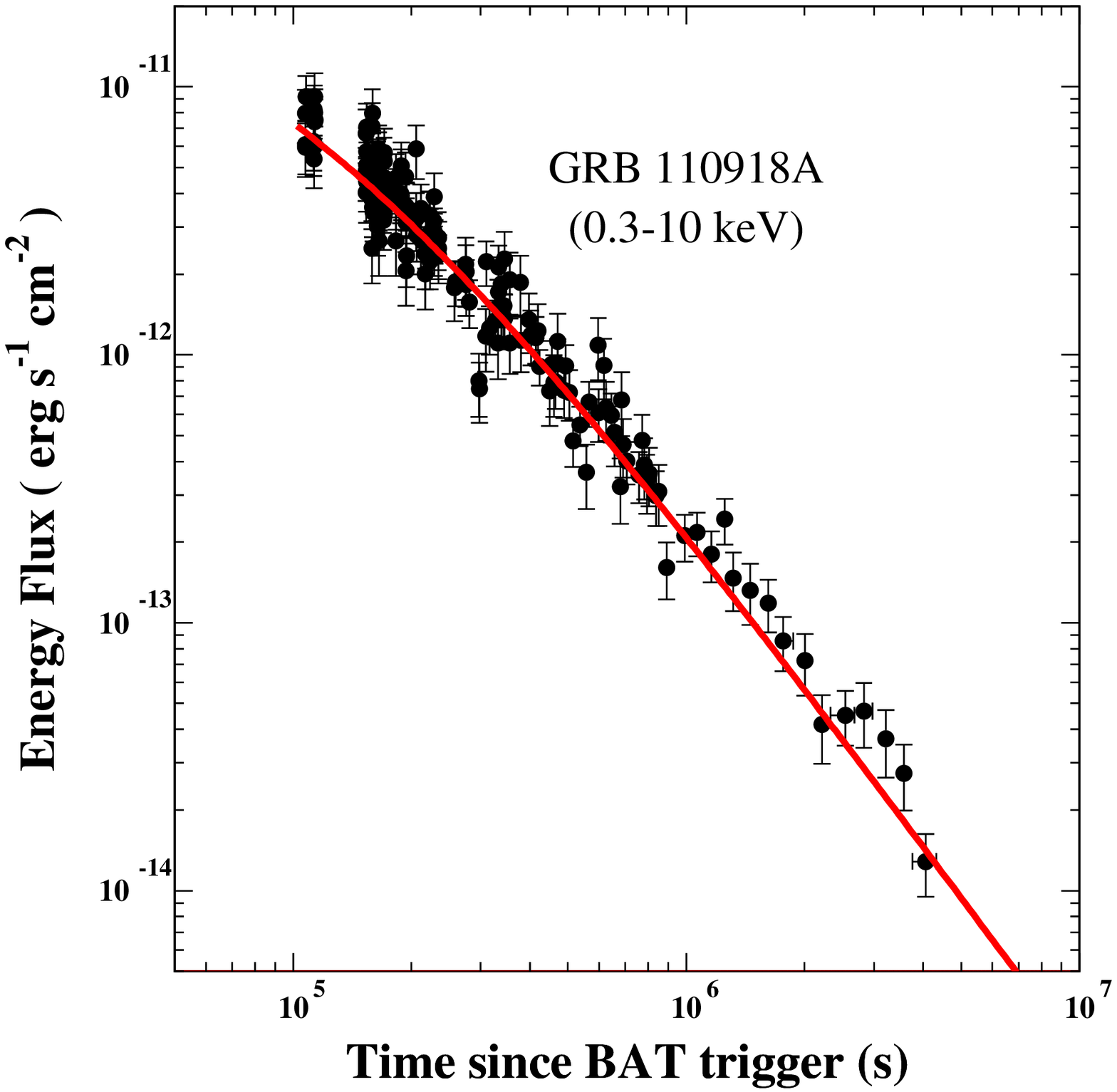,width=8.cm,height=8.cm} 
\epsfig{file=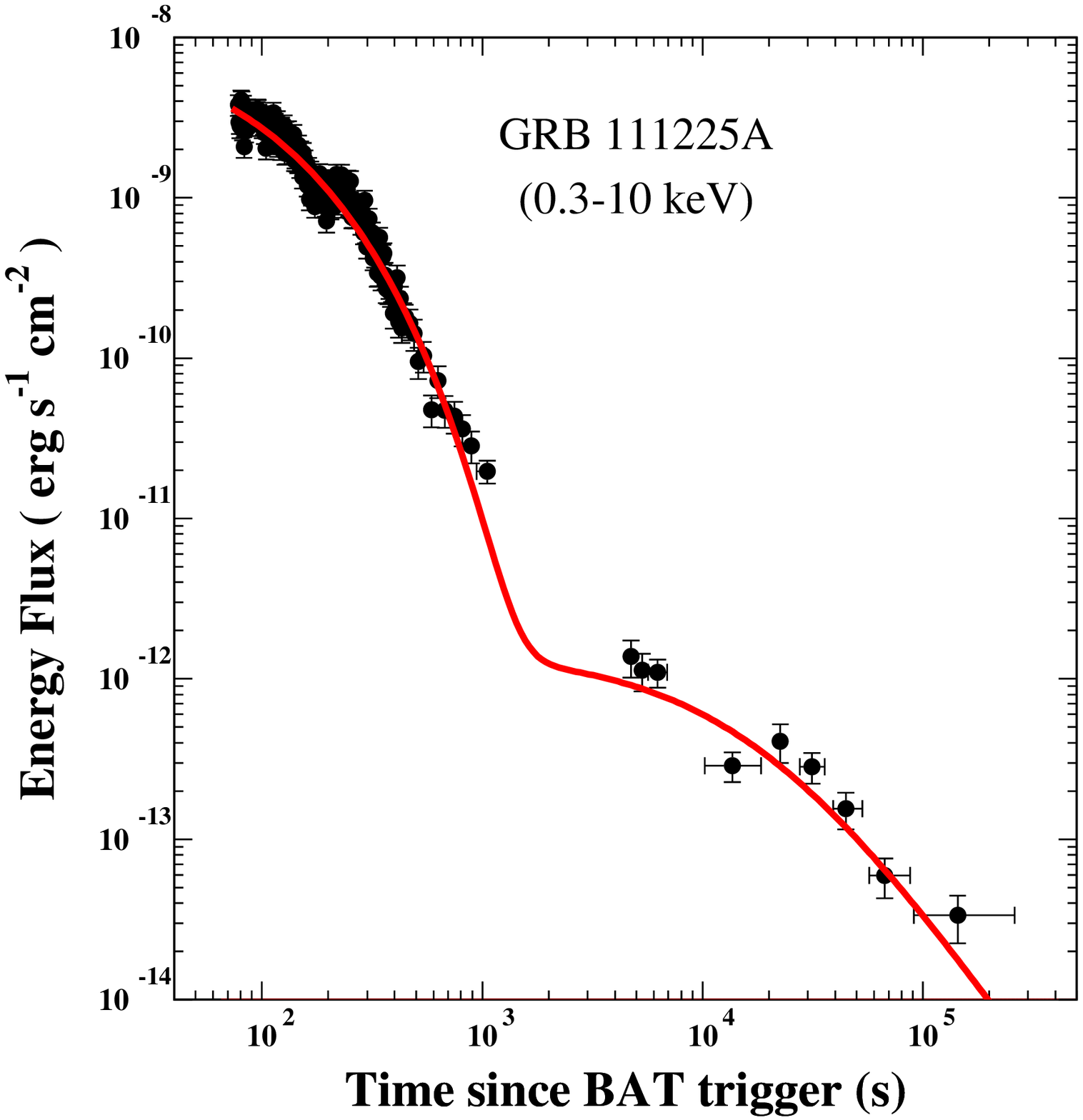,width=8.cm,height=8.cm}
}} 
\caption{The X-ray light curve of the nearby ($0.15<z<1$) 
SN-less GRBs 091127, 100418A, 100508A, 110106B, 110918A  
and 111225A reported in the Swift-XRT GRB light curve 
repository (Evans et al. 2007,2009)
and their best fit light curve for a jet contribution taken over 
by MSP powered emission, as given by Eq.(10) with the parameters 
listed in Table 3.}
\label{Fig6}  
\end{figure}

\begin{figure}[]
\centering
\vspace{-1cm}
\vbox{
\hbox{
\epsfig{file=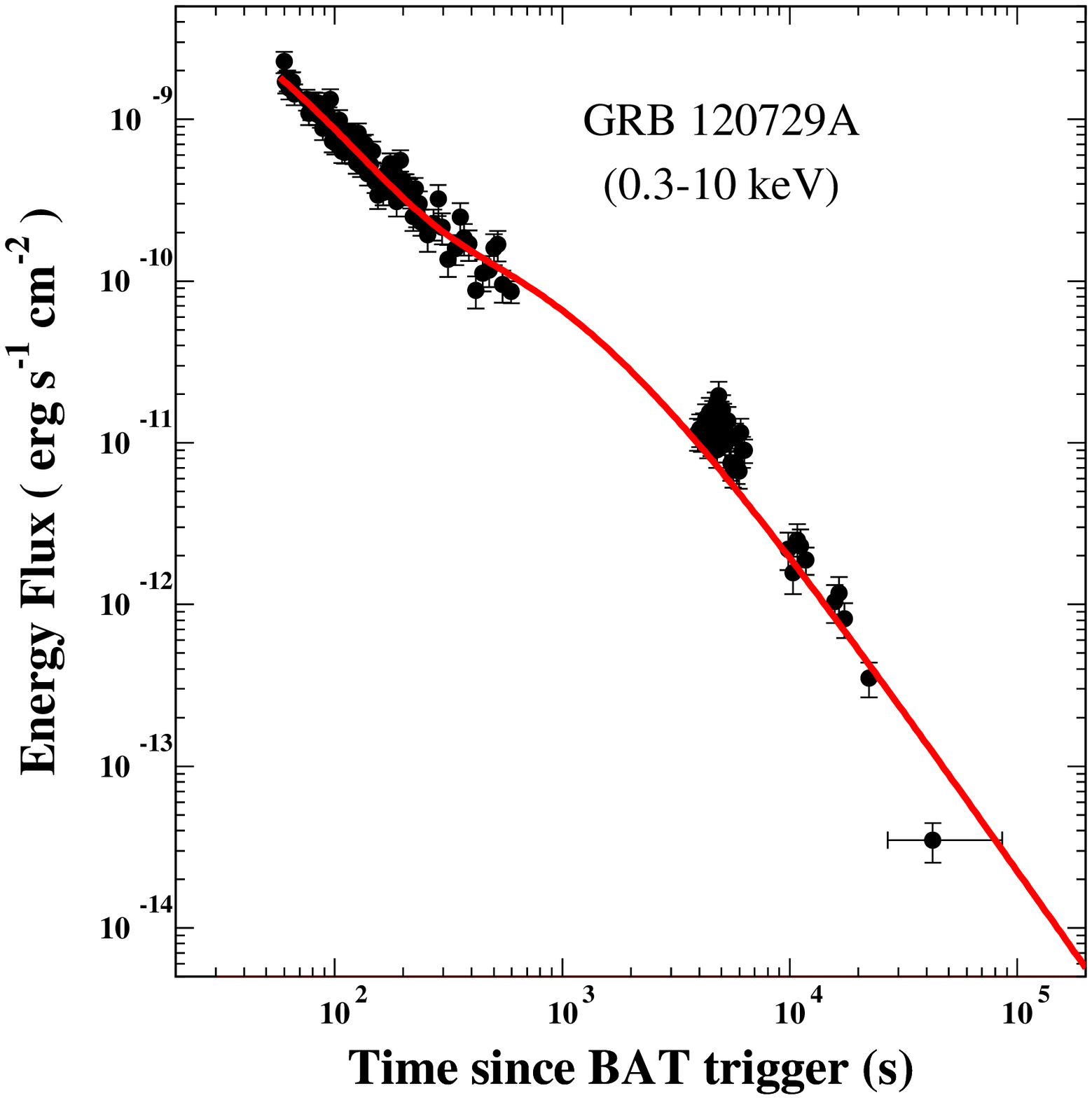,width=8.cm,height=8.cm}
\epsfig{file=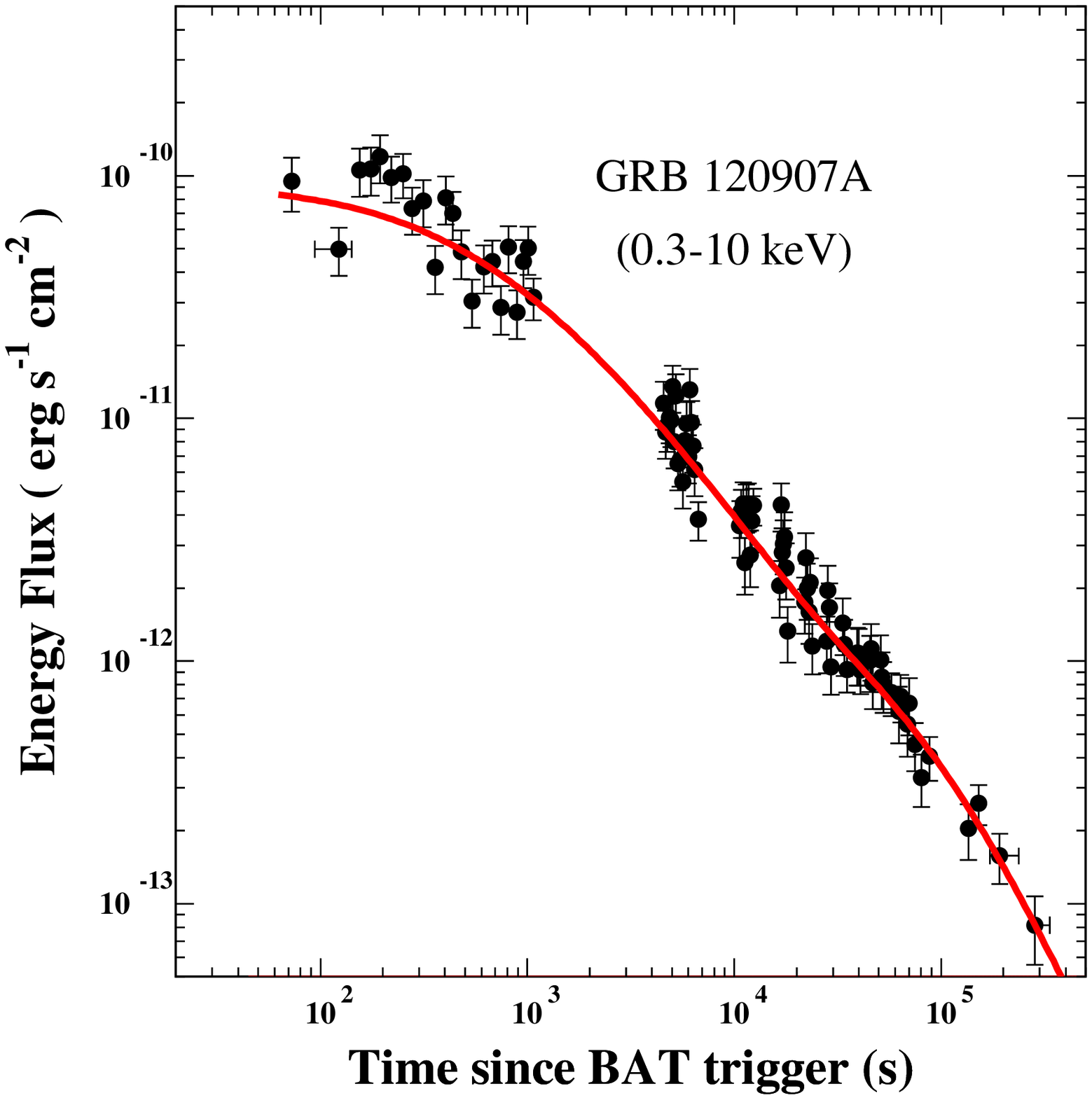,width=8.cm,height=8.cm}
}}
\vbox{
\hbox{
\epsfig{file=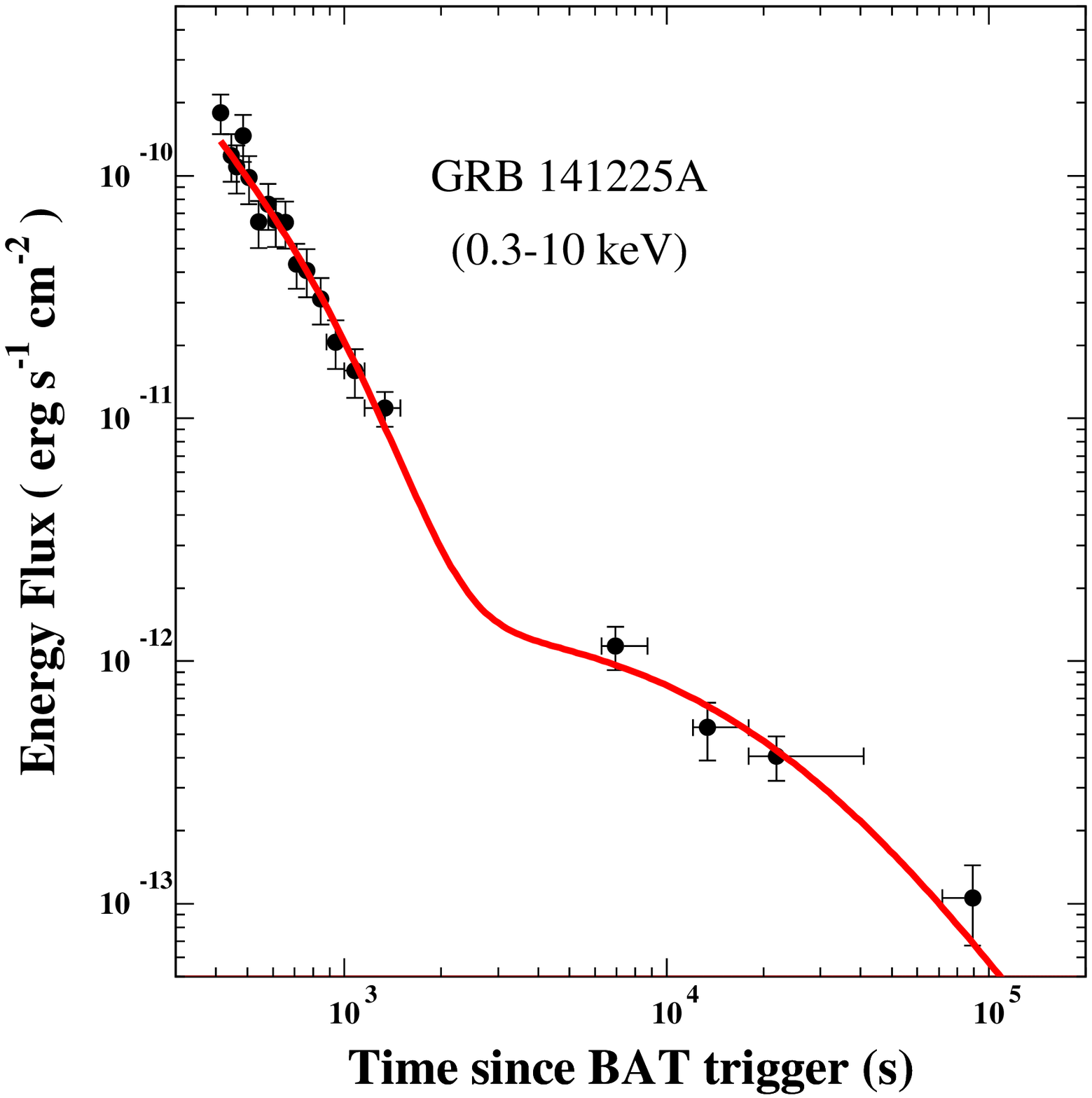,width=8.cm,height=8.cm}
\epsfig{file=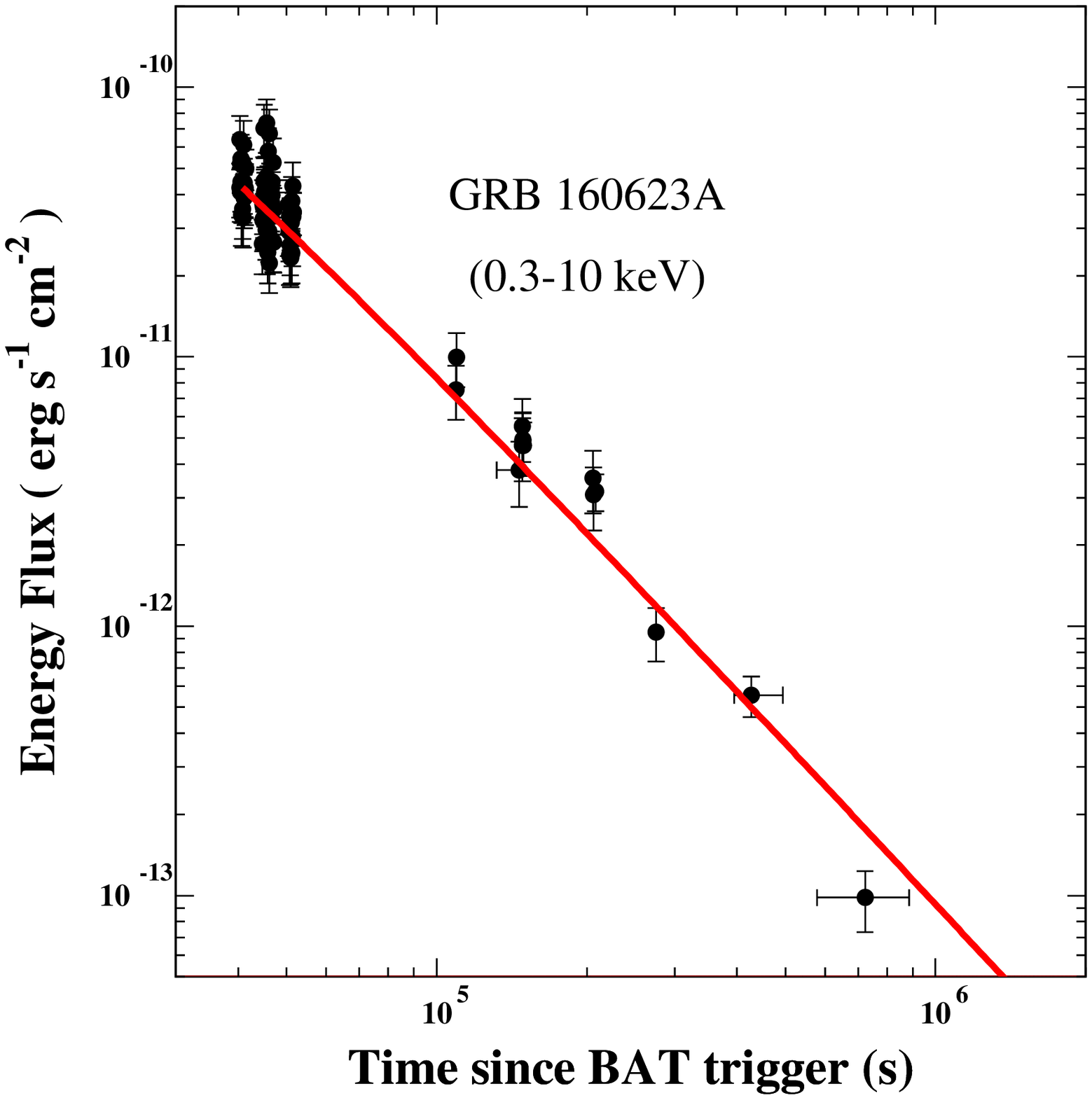,width=8.cm,height=8.cm}
}}
\vbox{
\hbox{

\epsfig{file=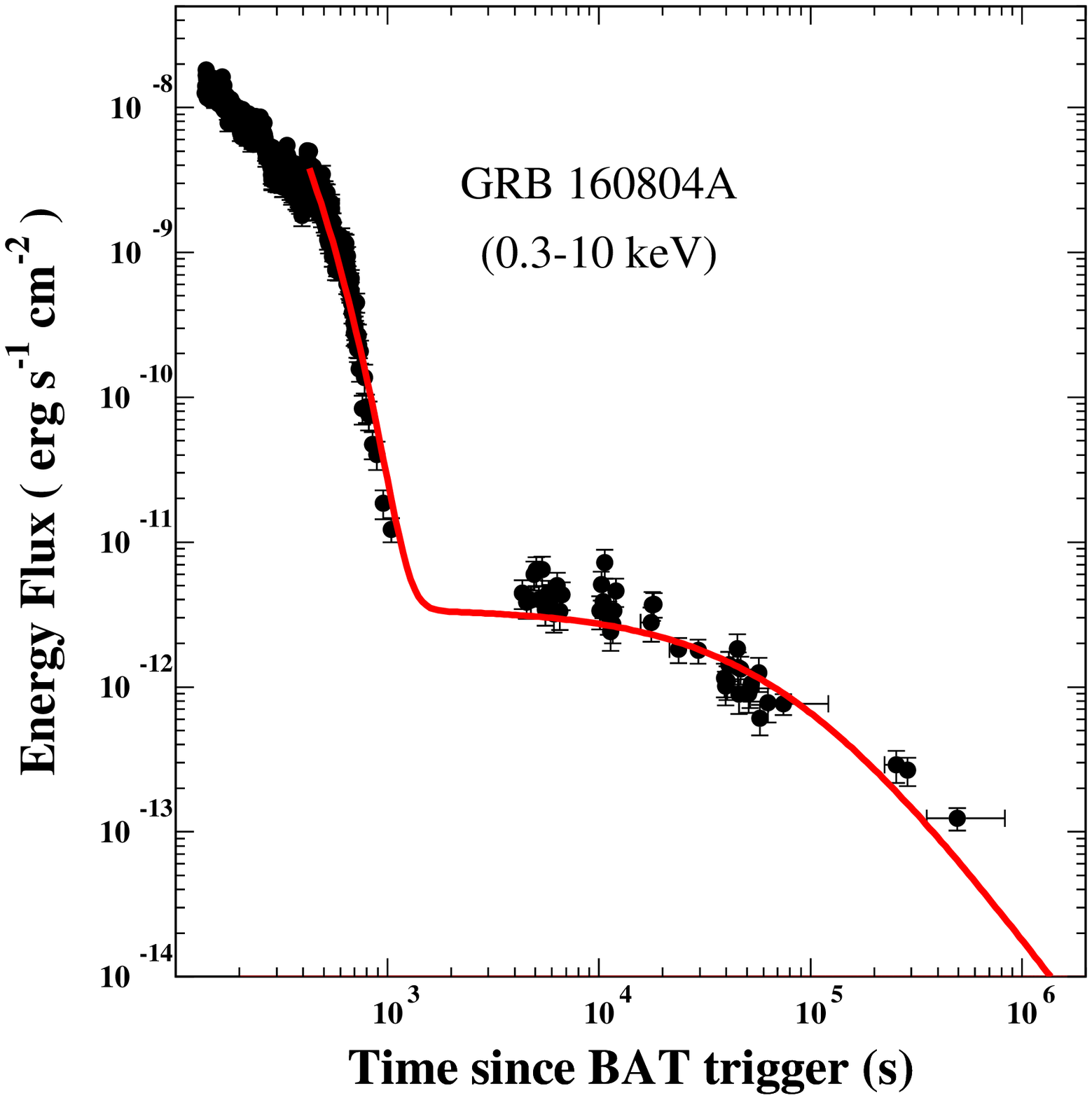,width=8.cm,height=8.cm}
\epsfig{file=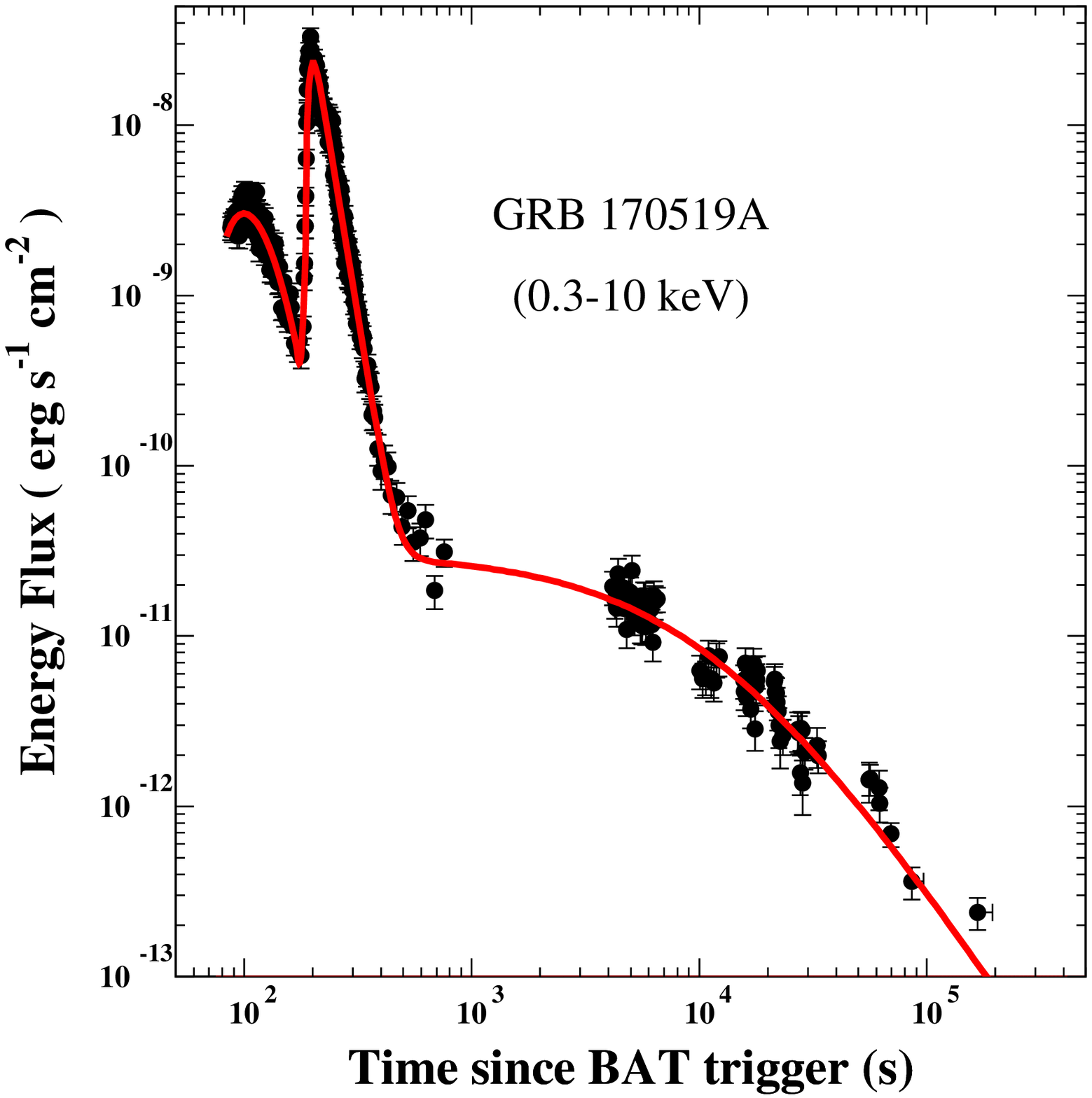,width=8.cm,height=8.cm}
}}
\caption{The X-ray light curve of the nearby ($0.15<z<1$) 
SN-less GRBs120729A, 120907A, 141225A, 160623A, 160804A 
and 170519A reported in the Swift-XRT GRB light curve 
repository (Evans et al. 2007,2009) and their best fit 
light curve for a jet contribution taken over by 
MSP-powered emission, as given by Eq.(10)  with the parameters
listed in Table 3.}
\label{Fig7}
\end{figure}

\begin{figure}[]
\centering
\vspace{-1cm}
\vbox{
\hbox{
\epsfig{file=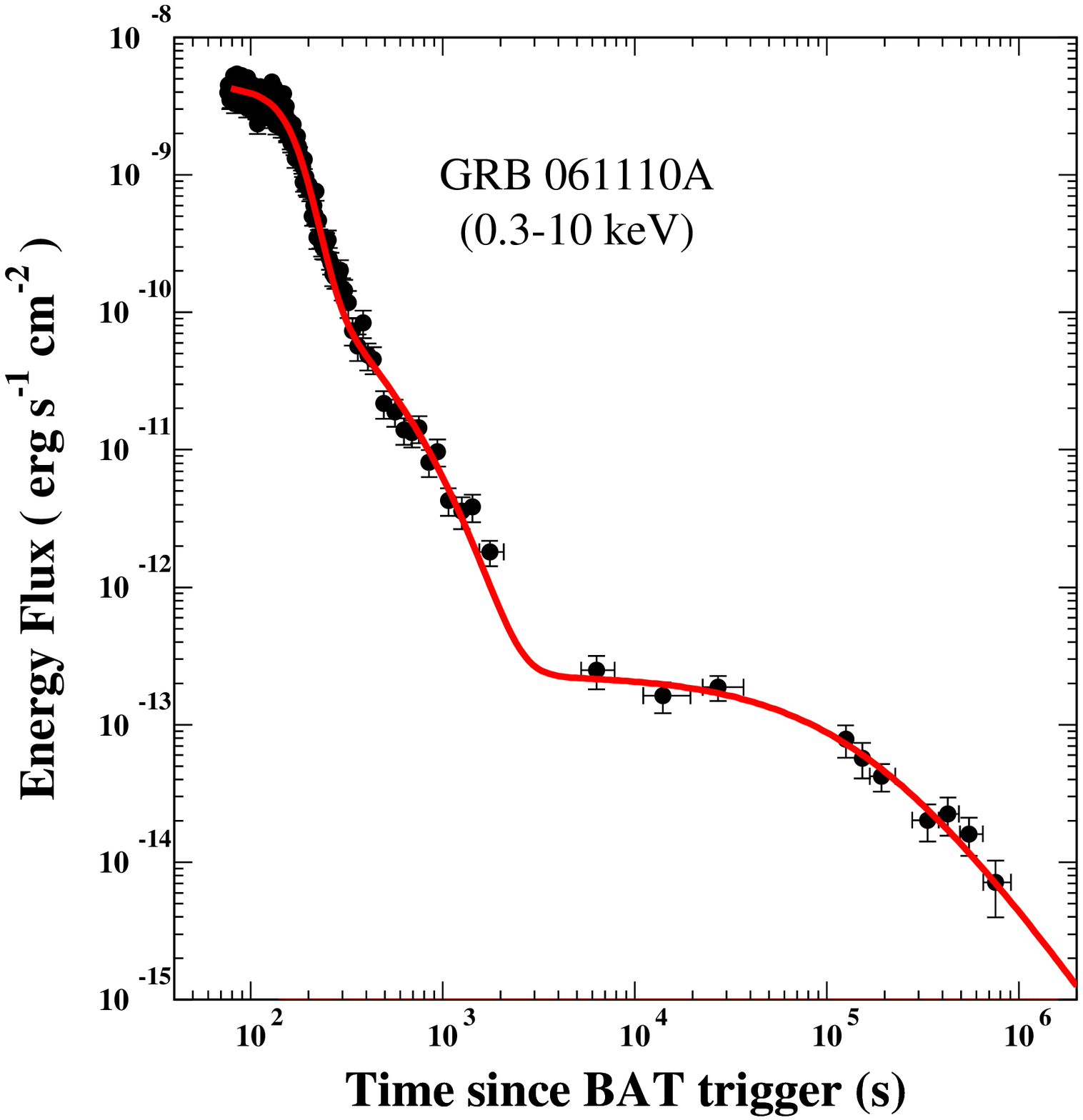,width=8.cm,height=8.cm}
\epsfig{file=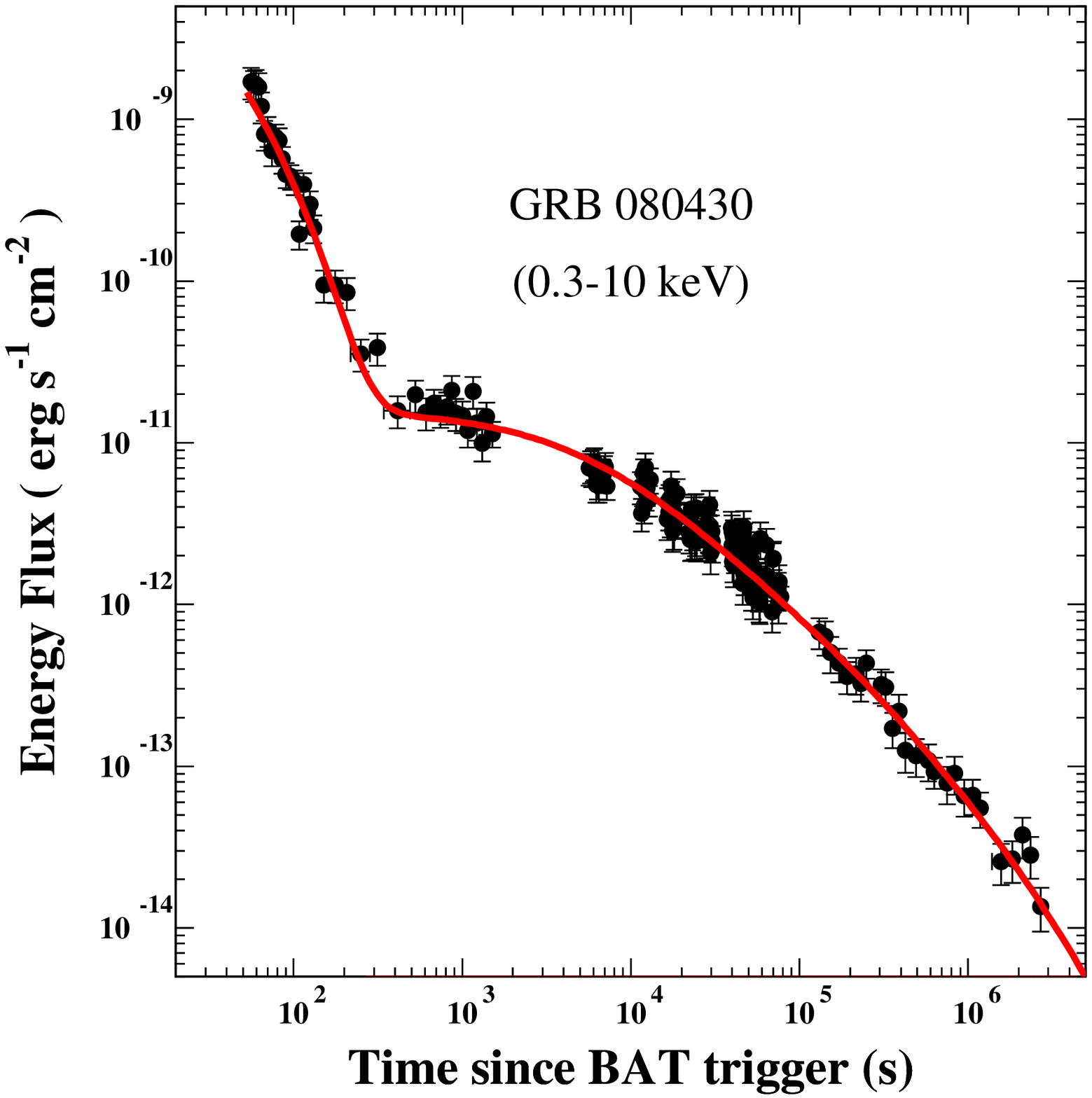,width=8.cm,height=8.cm}
}}
\vbox{
\hbox{
\epsfig{file=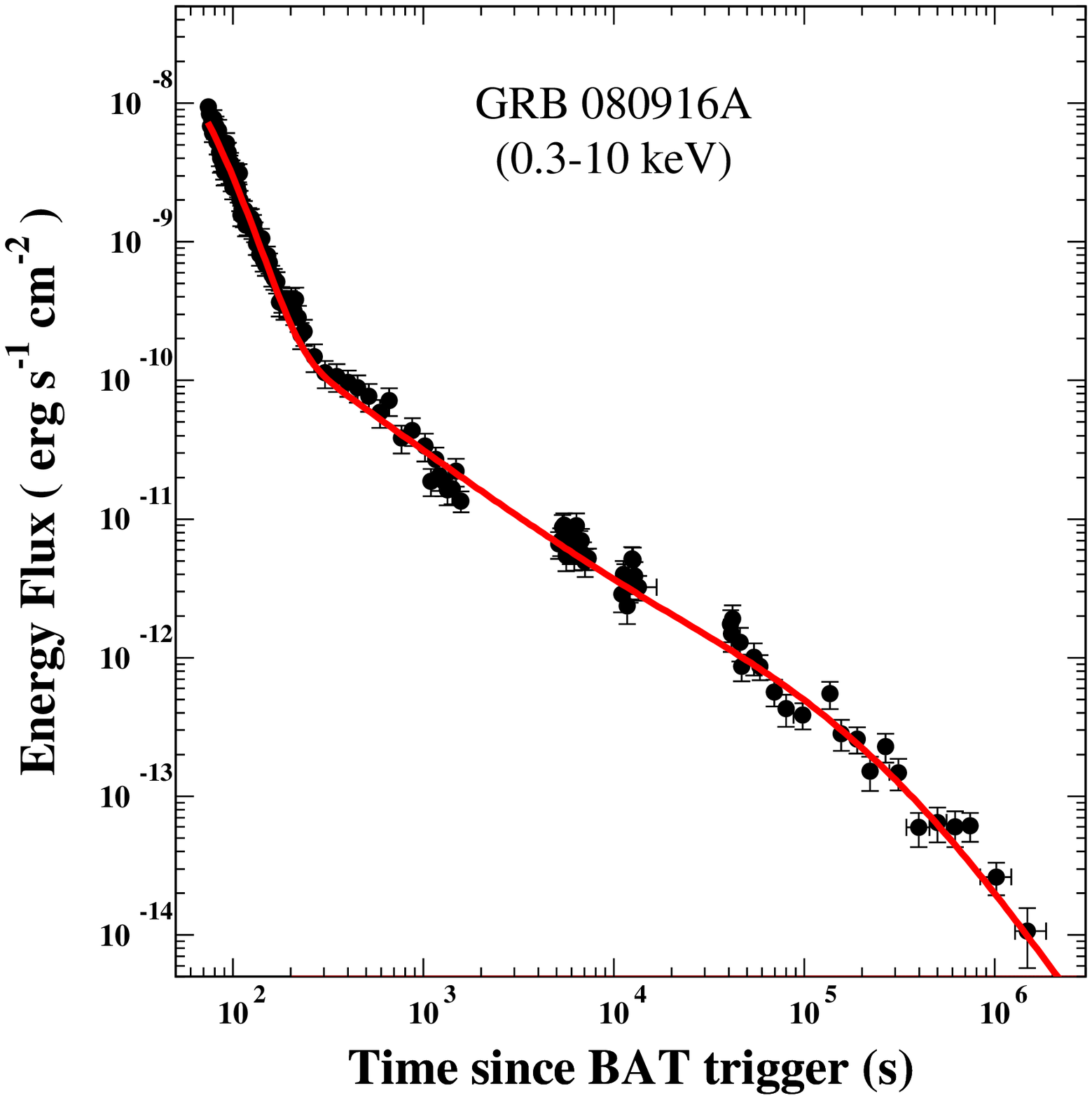,width=8.cm,height=8.cm}
\epsfig{file=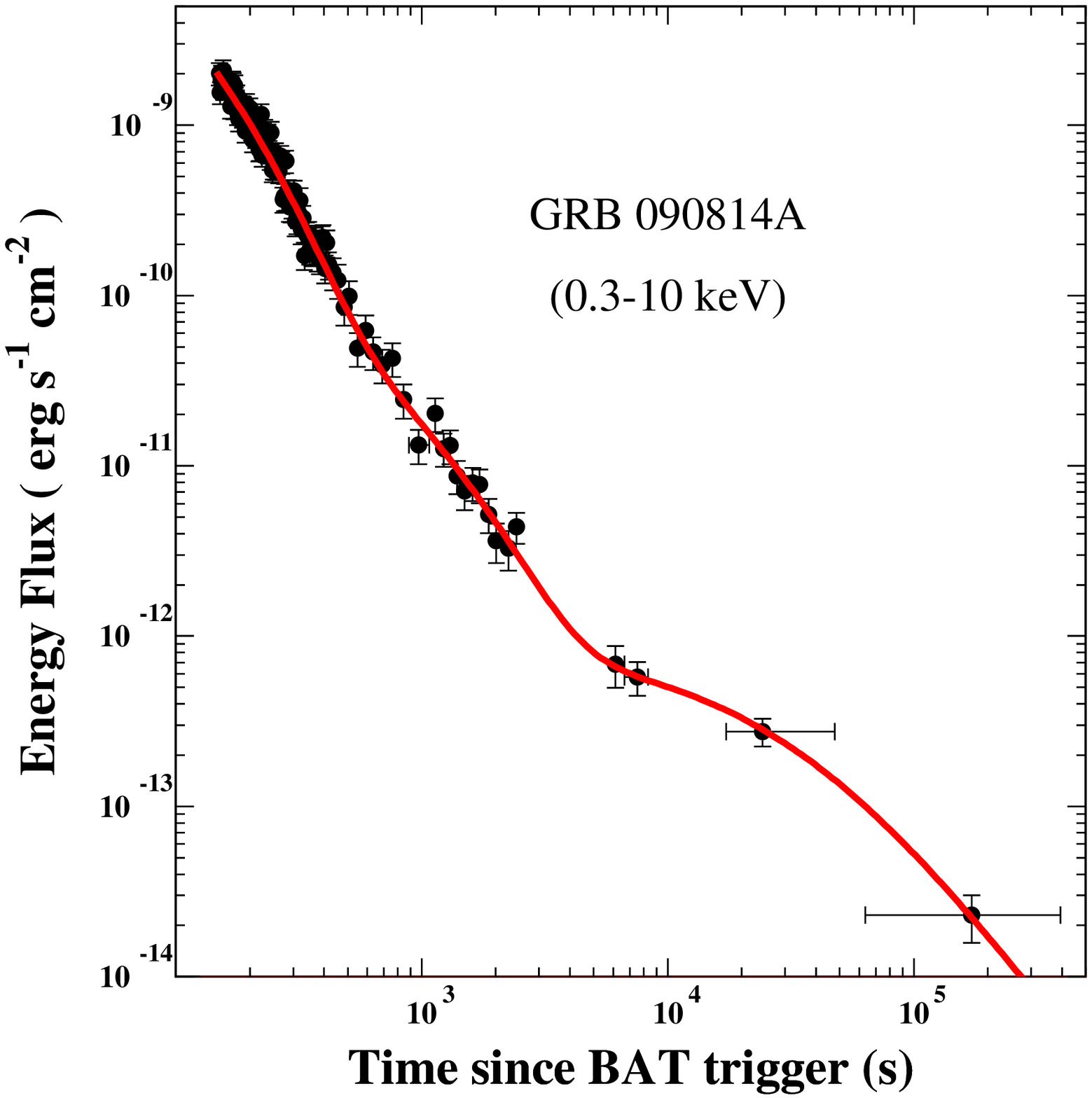,width=8.cm,height=8.cm}
}}
\vbox{
\hbox{
\epsfig{file=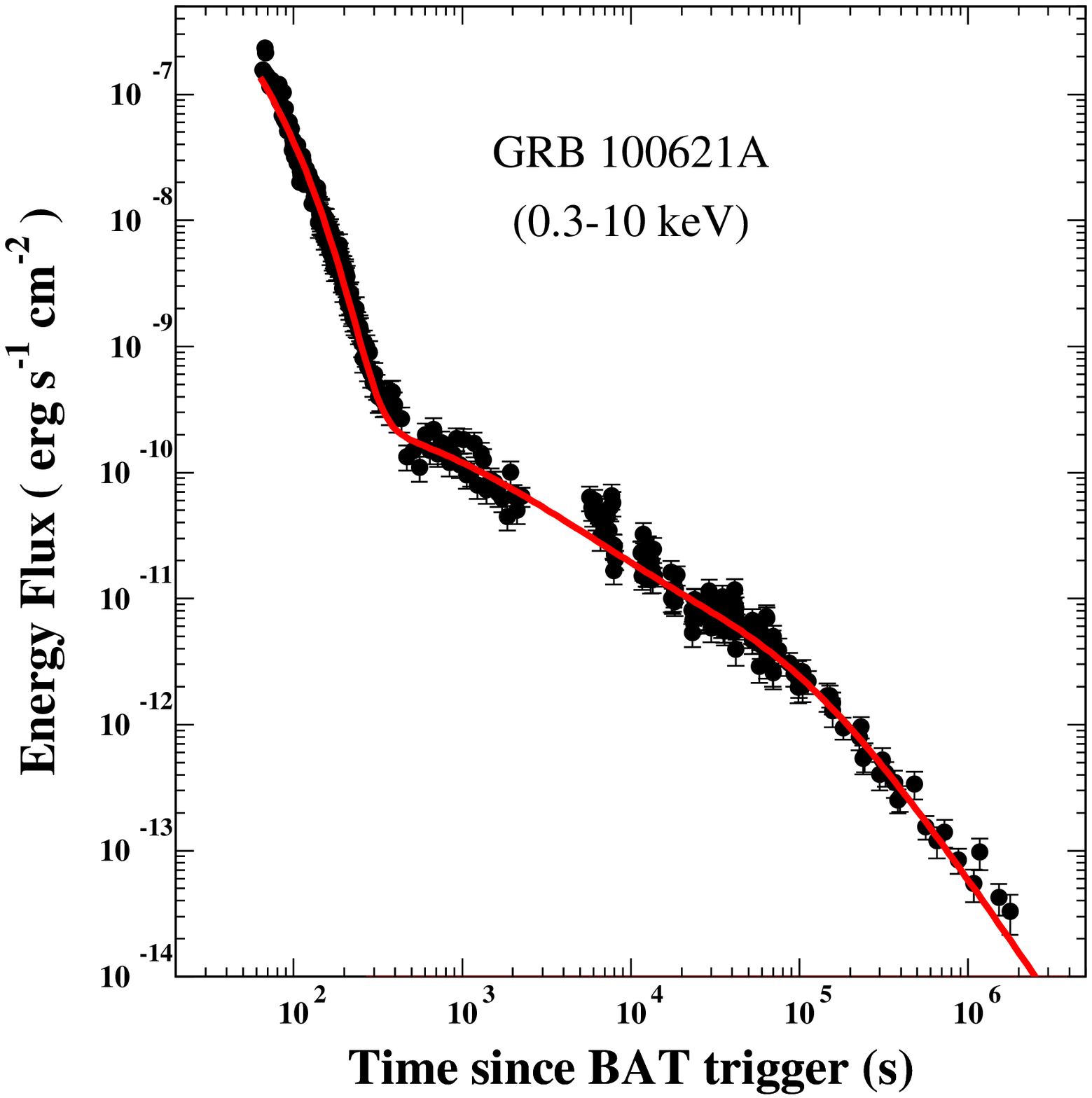,width=8.cm,height=8.cm}
\epsfig{file=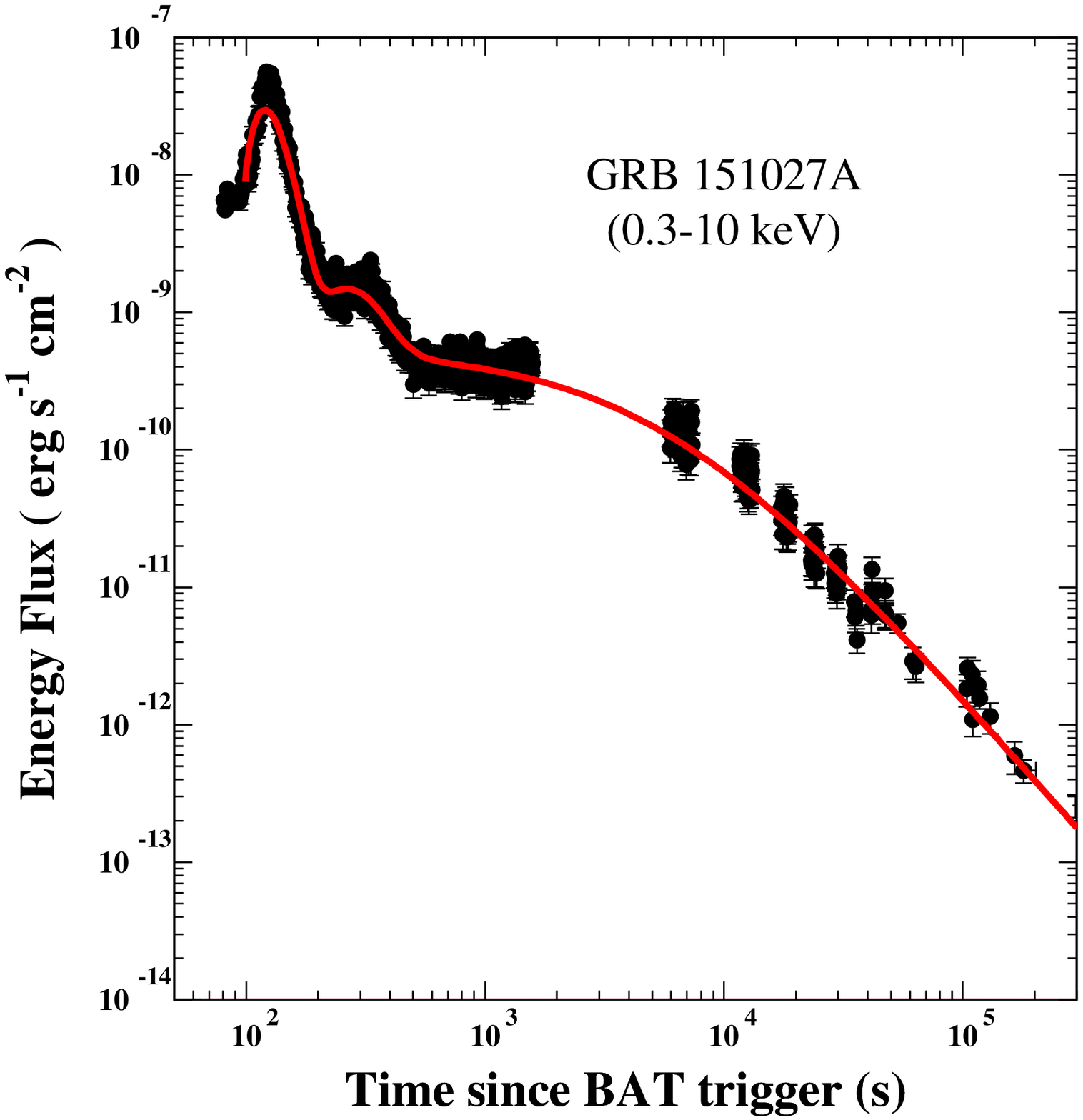,width=8.cm,height=8.cm}
}}
\caption{The X-ray light curve of the nearby ($0.15<z<1$)
SN-less GRBs 061110A, 080430, 080916A, 090814A, 100621A, 
and  151027A,   reported in the 
Swift-XRT GRB light curve repository (Evans et al. 2007,2009),
and their best fit light curve for a jet  contribution taken over
by MSP powered emission, as given by Eq.(4) plus Eq.(7)  with 
the parameters listed in Table 3.}
\label{Fig8}
\end{figure}

\begin{figure}[]
\centering   
\vspace{-1cm}
\vbox{
\hbox{
\epsfig{file=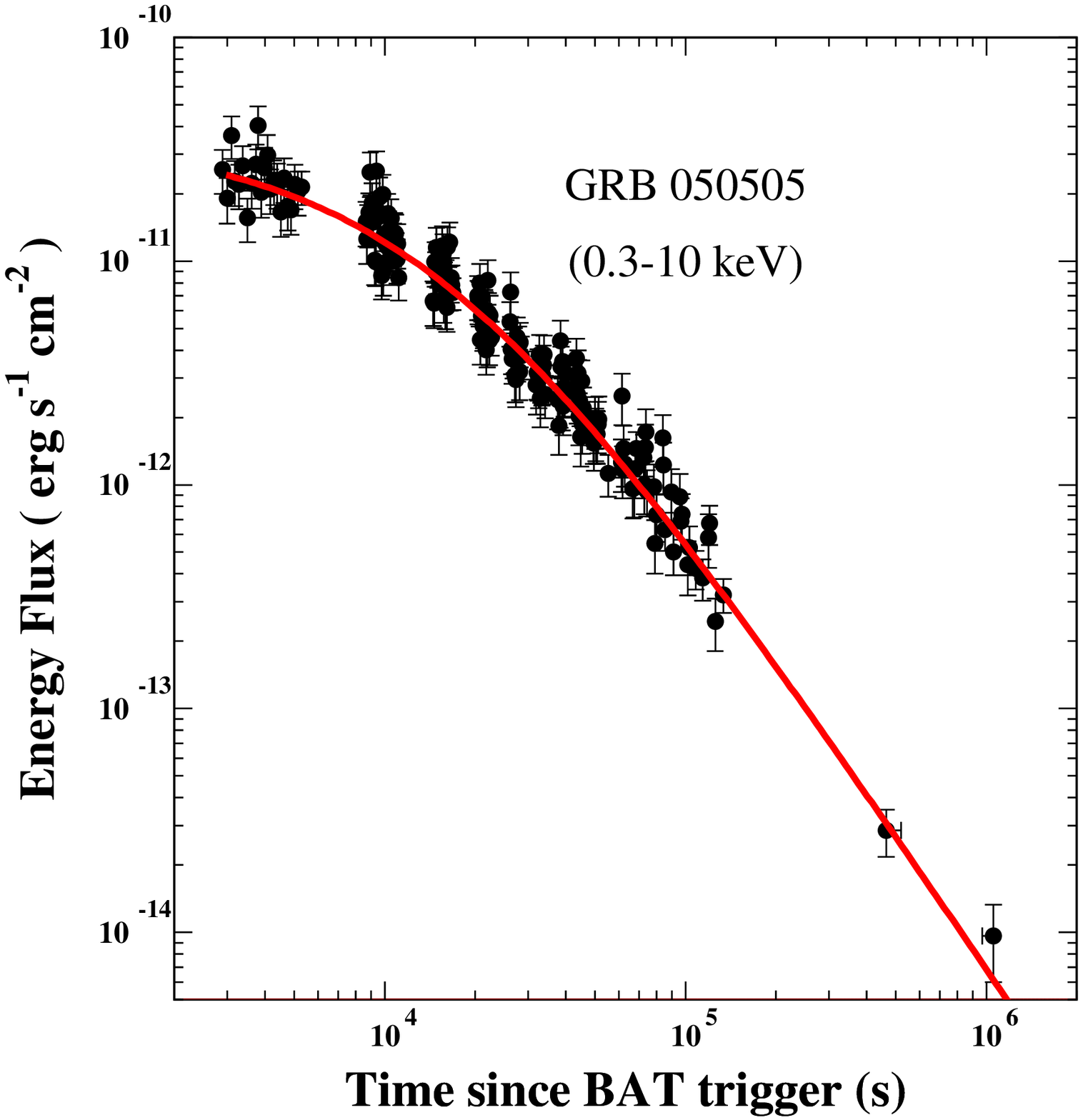,width=8.cm,height=8.cm}
\epsfig{file=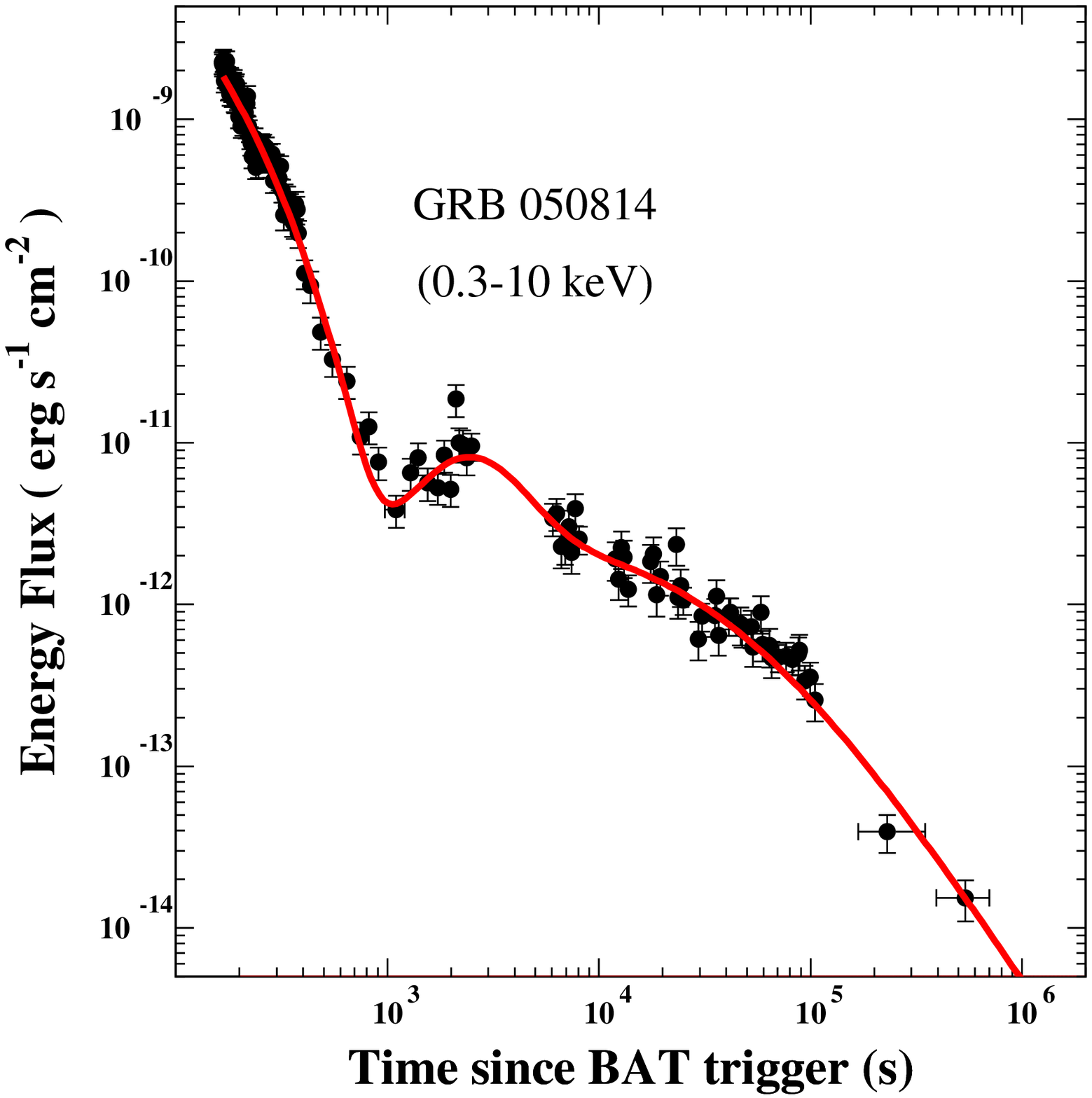,width=8.cm,height=8.cm}
}}
\vbox{
\hbox{
\epsfig{file=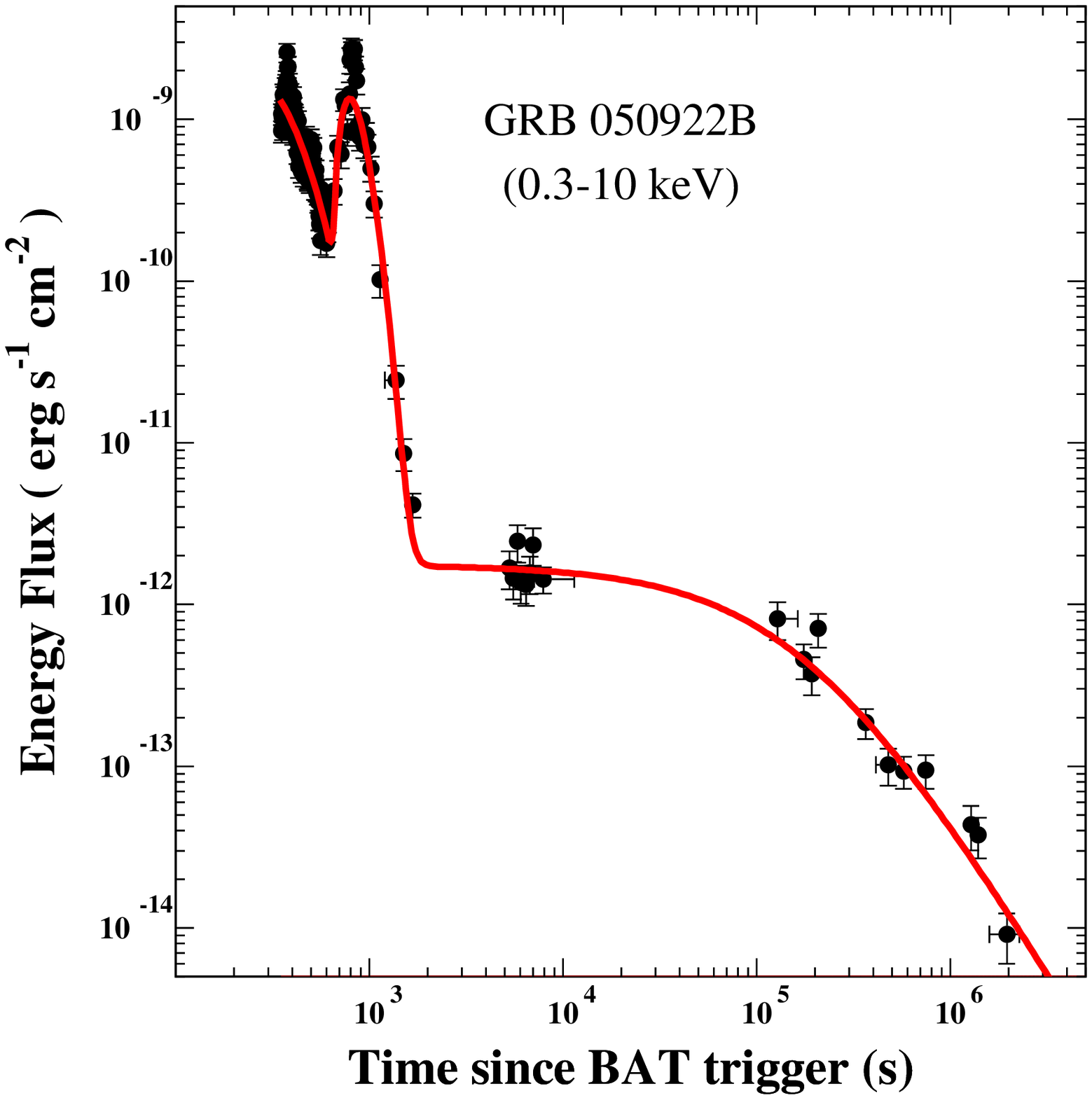,width=8.cm,height=8.cm}
\epsfig{file=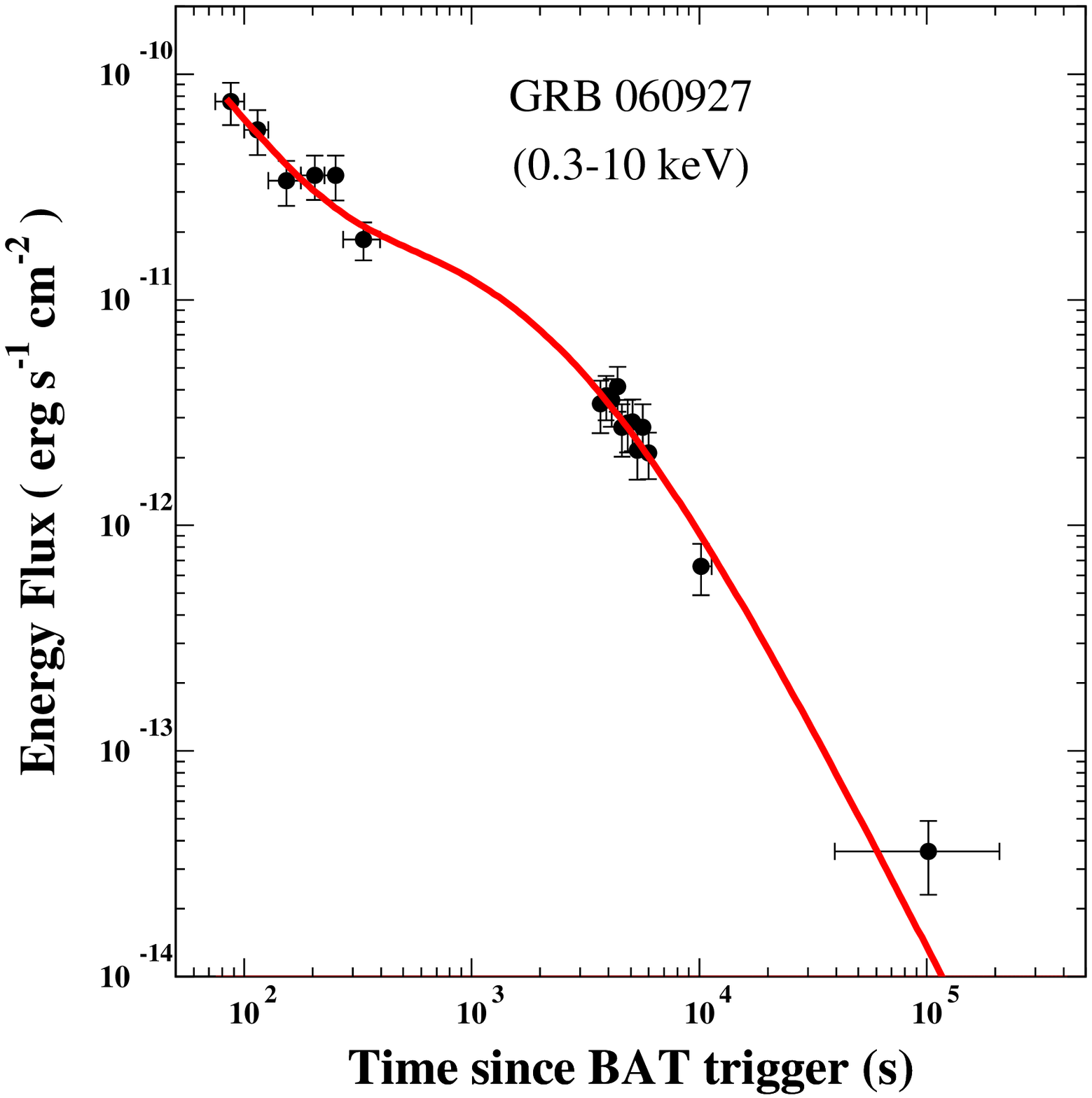,width=8.cm,height=8.cm}
}}
\vbox{
\hbox{
\epsfig{file=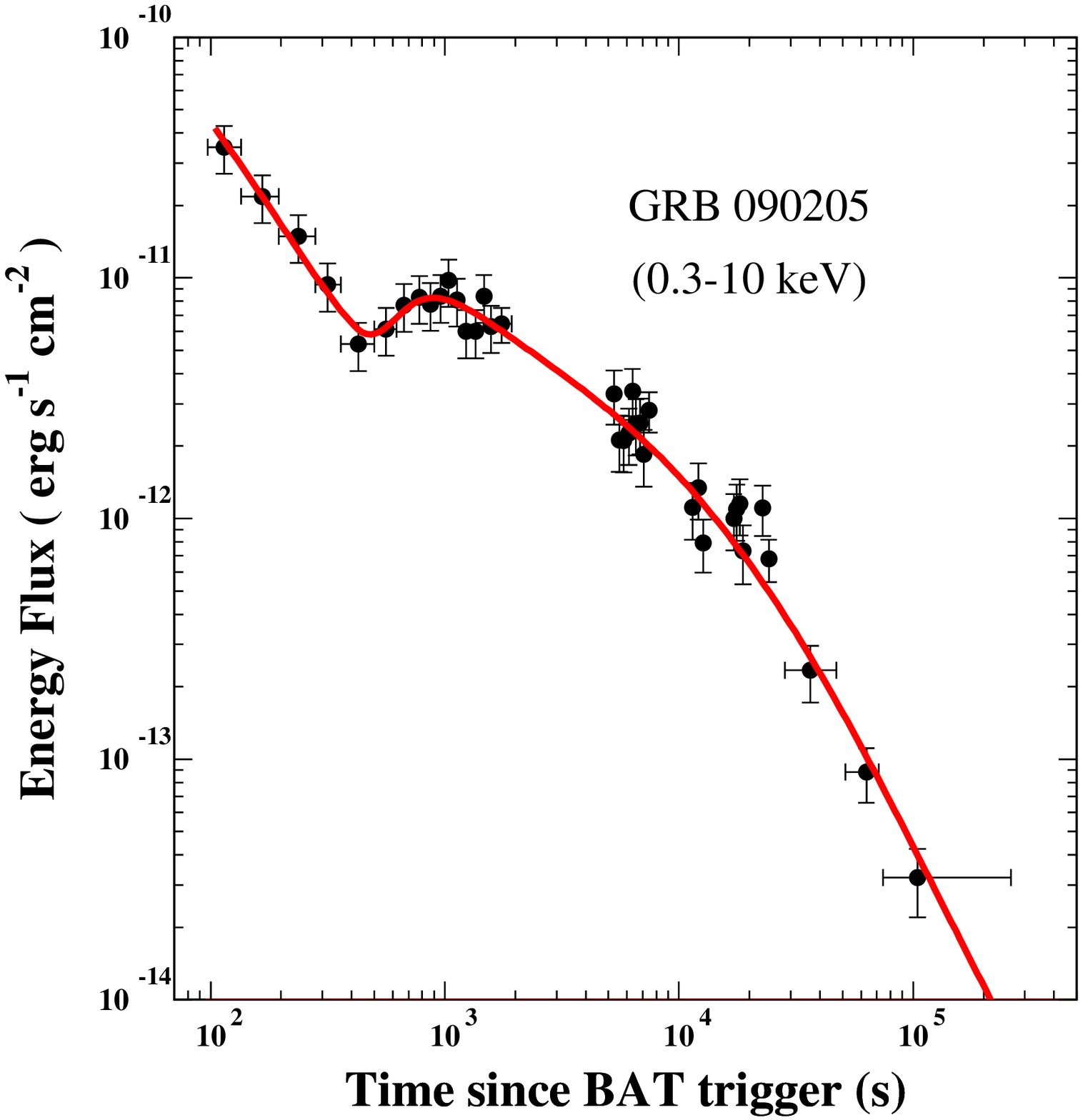,width=8.cm,height=8.cm} 
\epsfig{file=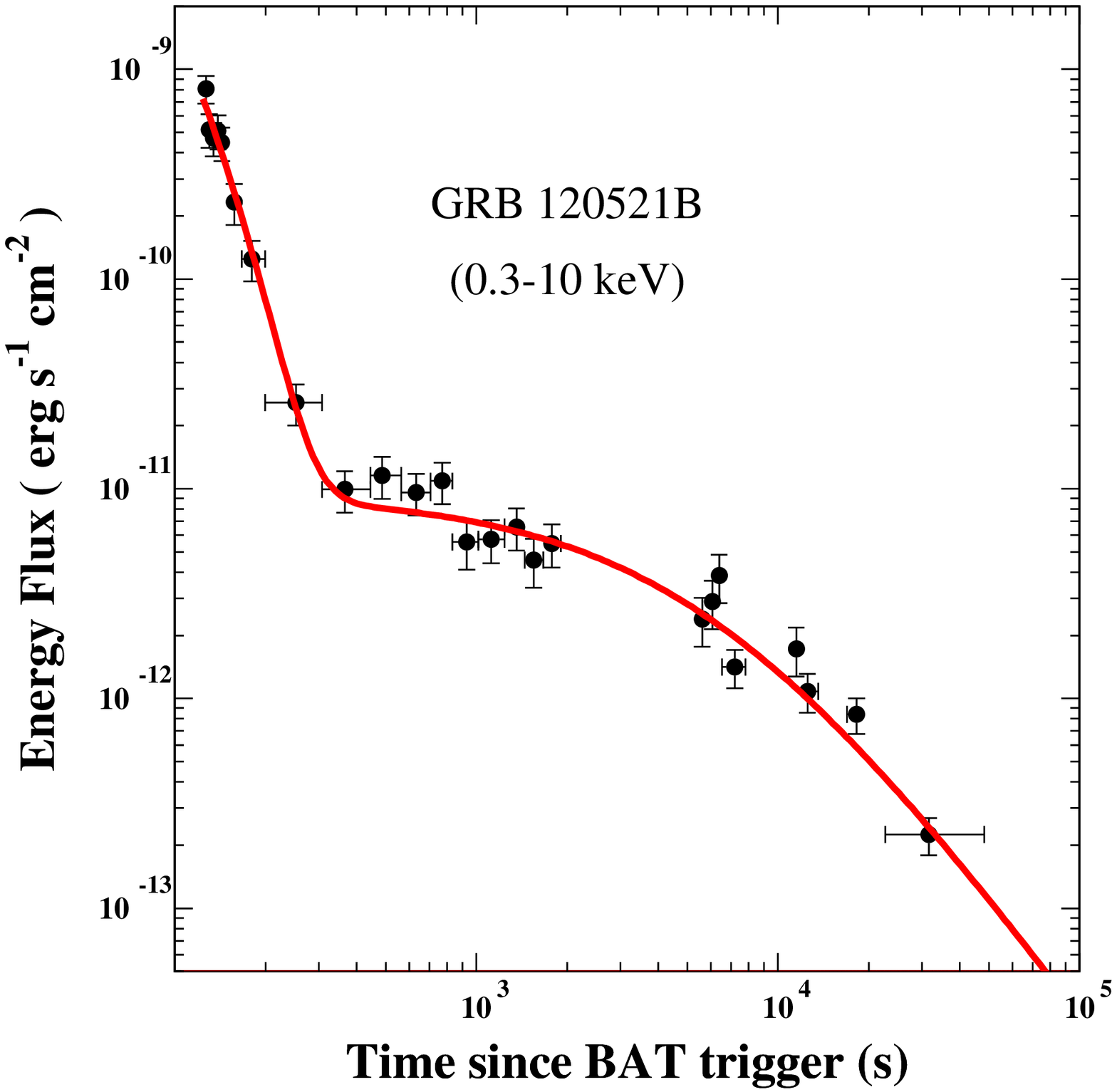,width=8.cm,height=8.cm}
}}
\caption{The X-ray light curve of the SN-less GRBs
050505, 050814, 050922B, 060927, 090205, and 120521B with $z>4$
reported in the Swift-XRT GRB light curve repository
(Evans et al. 2007,2009)
and their best fit light curve for a jet contribution taken over
by MSP powered emission, as given by Eq.(10) or by Eq.(4) plus Eq.(7)  
with the parameters listed in Table 4.}
\label{Fig9}
\end{figure} 

\begin{figure}[]
\centering   
\vbox{
\hbox{
\epsfig{file=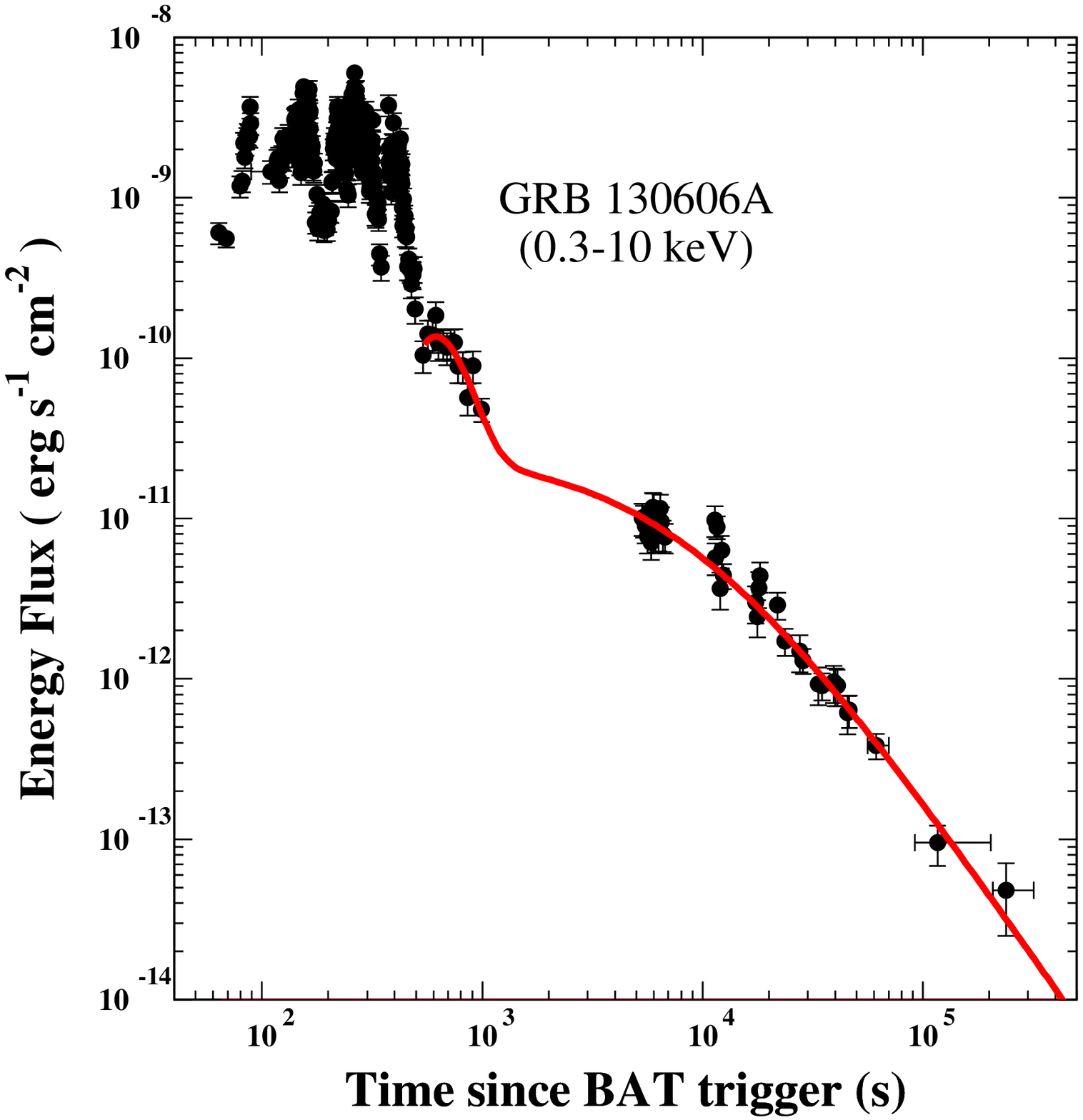,width=8.cm,height=8.cm}
\epsfig{file=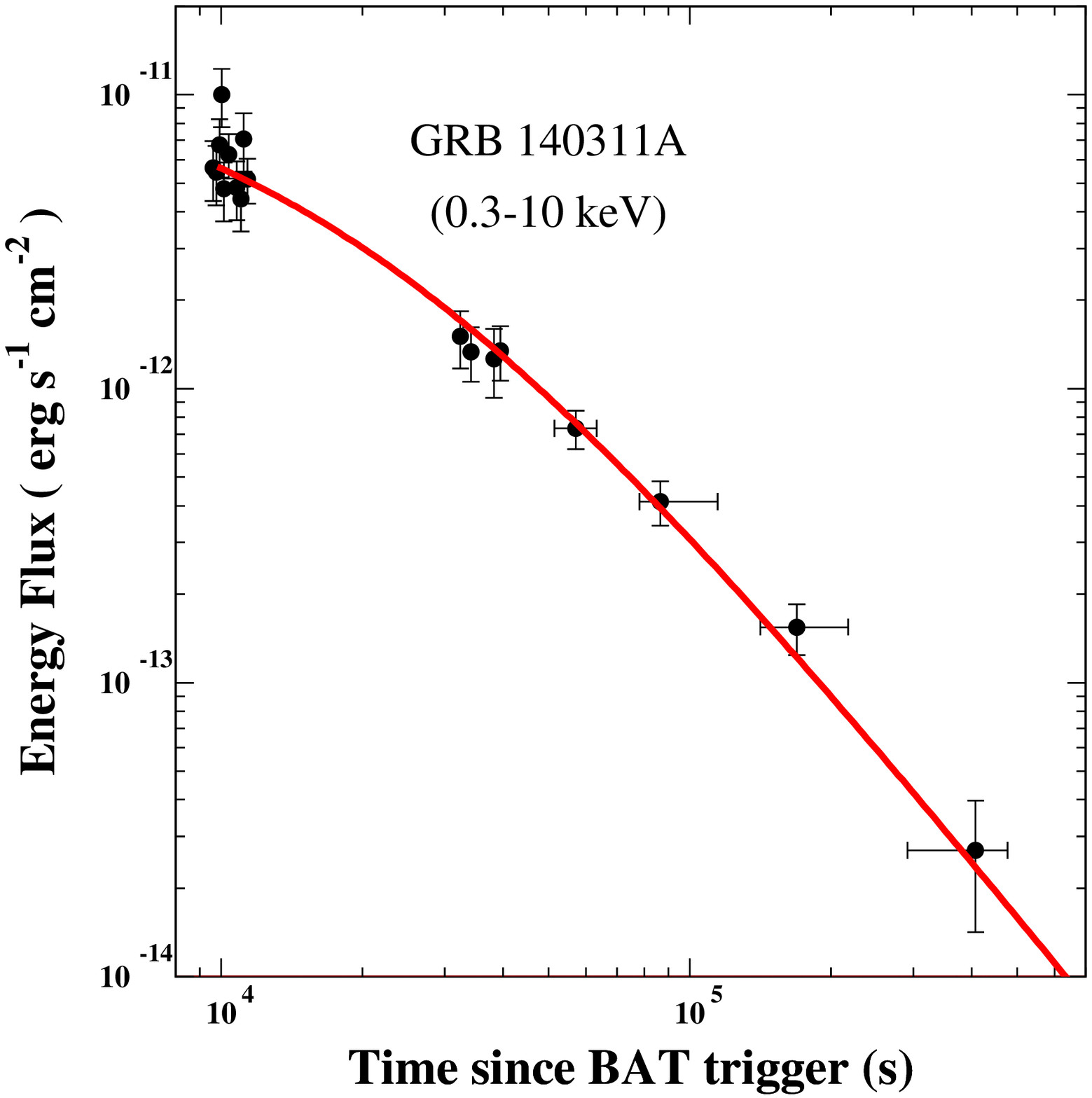,width=8.cm,height=8.cm}
}}
\vbox{
\hbox{
\epsfig{file=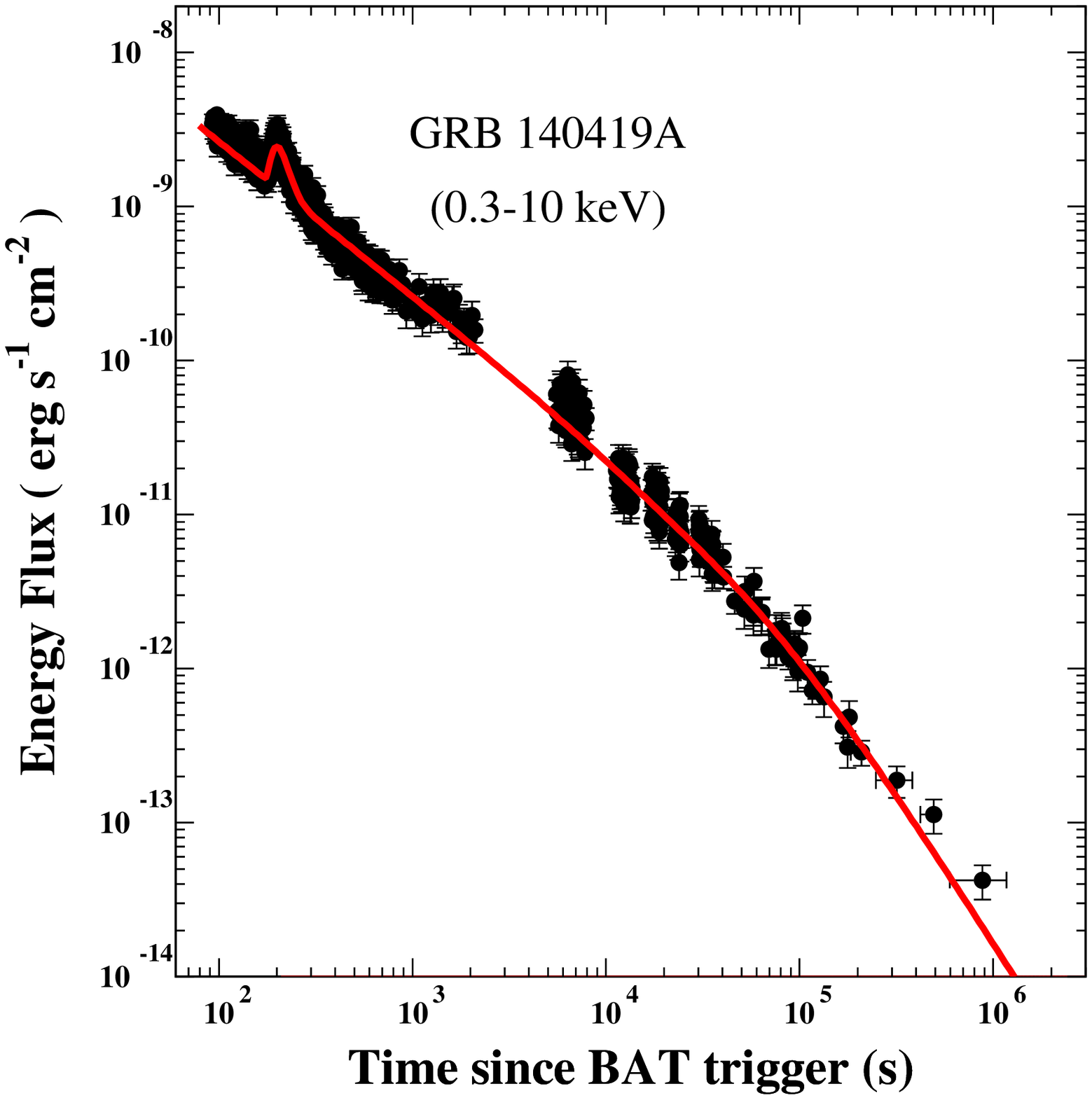,width=8.cm,height=8.cm}
\epsfig{file=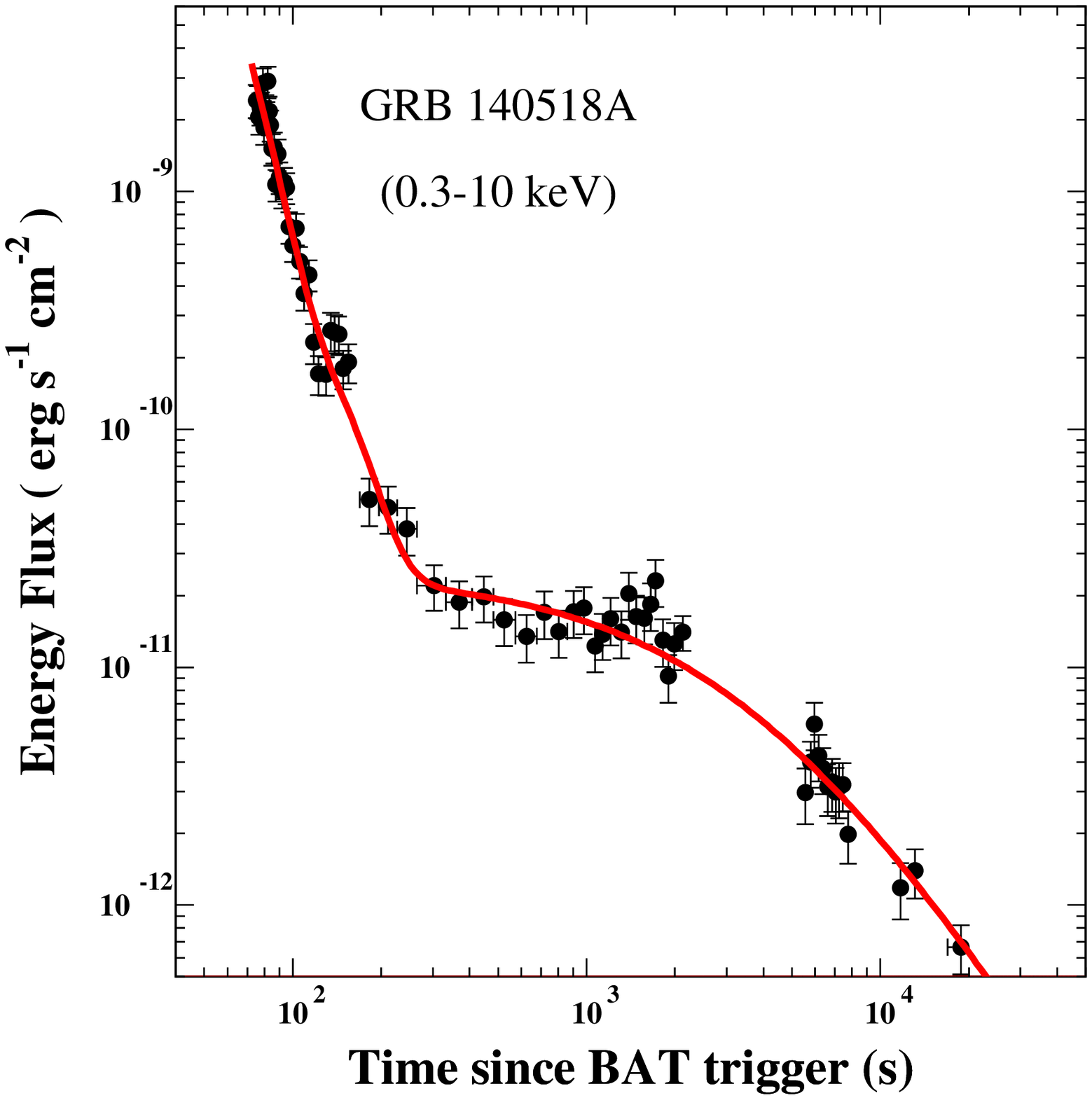,width=8.cm,height=8.cm}
}}
\vbox{
\hbox{
\epsfig{file=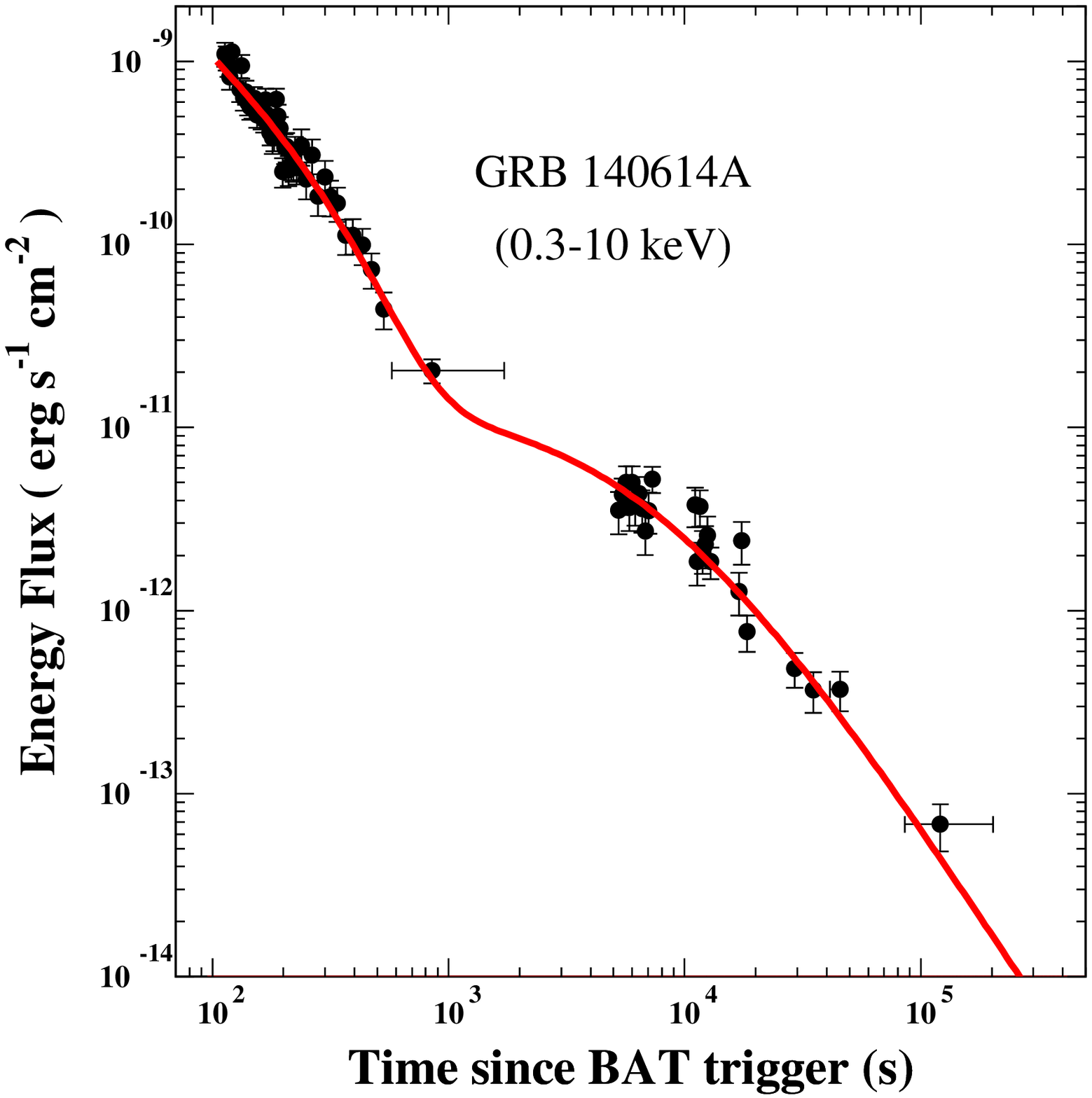,width=8.cm,height=8.cm}
\epsfig{file=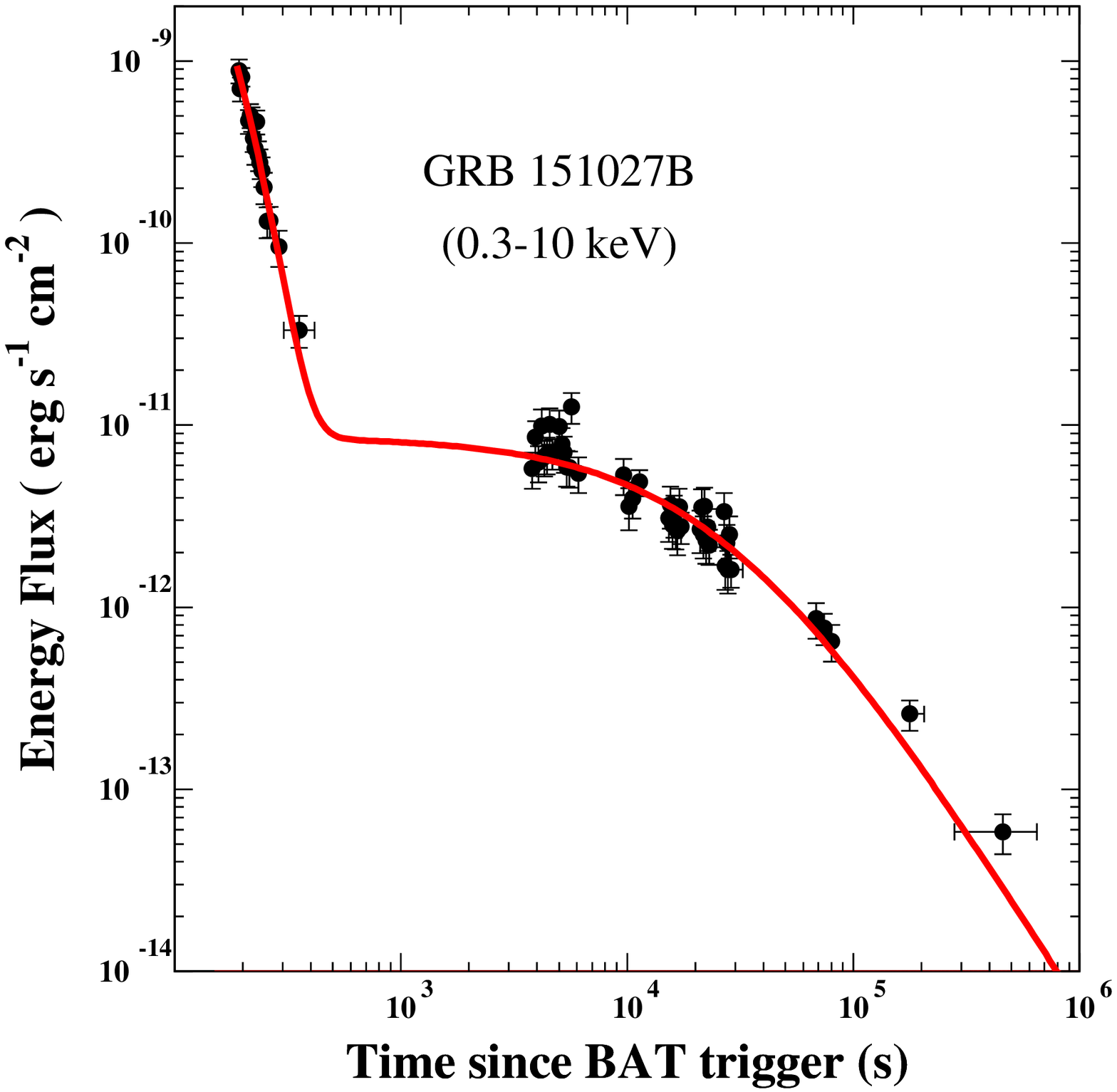,width=8.cm,height=8.cm}
}}
\caption{The X-ray light curve of the SN-less GRBs
130606A, 140311A, 140419A, 140518A, 140614A, and 151027B with $z>4$
reported in the Swift-XRT GRB light curve repository
(Evans et al. 2007,2009)
and their best fit light curve for a jet contribution taken over
by MSP powered emission, as given by Eq.(10) or by Eq.(4) plus Eq.(7) 
with the parameters listed in Table 4.}
\label{Fig10}
\end{figure}

\begin{table*}
\caption{The values of the parameters of Eq.(10) obtained 
from a best fit to the well-sampled light curves of the X-ray 
afterglows of the 4 nearest SN-less GRBs ($z<0.15$) shown in Figures 1-4. Also
listed are the values of the periods  of
the MSPs at birth, as estimated from the best fit valued of
$F_{ps}$ and $t_b$.}
\centering
\begin{tabular}{l l l l l l l l l l l}
\hline
\hline
~~SHB~~~&~~z~~~&~~~$F_i$~~~~&~a~&~$\tau$~~~~&~~$F_{ps}$~~&~$t_b$&$\chi^2/dof$&$P_0/\eta^{1/2}$\\
        &     &$[erg/s\,cm^2\,]$&        &   ~[s]~  &$[erg/s\,cm^2]$~& ~[s]~&   & [ms]\\
\hline
051109B  &0.080 ?&  1.79E-7~ & [0] & ~~~15  & 6.21E-12 & ~~5958 &~~2.84~& ~?~  \\
080517~  &0.089~ &  4.83E-8~ & [0] & ~~153  & 3.92E-13 & ~33400 &~~2.84~& 303  \\
060614~  &0.125~ &  1.64E-7~ & [0] & ~~111  & 1.09E-11 & ~49830 &~~1.39~& ~33  \\
050826A  &0.297~ &  7.57E-9~ & [0] & ~~159  & 2.30E-11 & ~10692&~~1.10~ &  20  \\

\hline
\end{tabular}
\end{table*}

\begin{table*}
\caption{The parameters  obtained from the best fit
light curves to   the well-sampled  X-ray afterglows of GRBs 
of known redshift  $0.15<z<1$, without
a confirmed association with SN.  
The 18  GRBs shown in Figures  5-7  were fit  with Eq.(10) and the last 6 GRBs  
were fit with Eq.(4) plus Eq.(7). 
Also listed are the periods  of
the MSPs at birth, as estimated from the best fit values of
$F_{ps}$ and $t_b$.  The index i is the pulse number}
\centering

\begin{tabular}{l l l l l l l l l l}
\hline
\hline
~~SHB~~~&~~z~~~&~~~$F_i$~~~~&~$t_i$~&$\tau_i$~~&~~$F_{ps}$~~&~$t_b$&$\chi^2/dof$&~$P_0/\eta^{1/2}$~\\
        &     &$[erg/s\,cm^2\,]$&        &   ~[s]~  &$[erg/s\,cm^2]$~& ~[s]~&   & [ms] \\
\hline
051022A  &0.809~ &           & 0 &        & 1.40E-10 & ~13007&~~1.48~& 2.63 \\
060123~  &0.56~~ &           & 0 &        & 2.15E-12 & 146432&~~0.85~& 9.25 \\
080710~  &0.845~ &           & 0 &        & 3.42E-11 & ~~6740&~~0.85~& 7.07 \\
090814A  &0.696  &  6.88E-9  & 0 & ~ 125  & 8.46E-13 & ~33274&~~0.90~& 71.2 \\
091003~  &0.8969 &           & 0 &        & 7.24E-12 & ~83890&~~1.19~& 4.10 \\
091018~  &0.971~ &  5.68E-10 & 0 & ~4528  & 7.82E-12 & ~27775&~~1.14~& 6.34 \\

091127~  &0.49~~ &  9.61E-9~ & 0 & 48480  & 9.38E-12 & 140532&~~1.09~& 5.17 \\
100418A  &0.6235 &  1.76E-8~ & 0 & ~~40   & 7.02E-13 & 235123&~~1.56~& 11.4 \\
100508A  &0.5201 &  1.00E-8~ & 0 & ~~163  & 2.18E-11 & ~12067&~~2.17~& 10.9 \\
110106B  &0.618~ &  2.55E-11 & 0 & ~~511  & 1.56E-11 & ~16076&~~1.42~& 9.32 \\
110918A  &0.982~ &           & 0 &        & 3.61E-11 & ~82033&~~1.24~& 1.69 \\
111225A  &0.297~ &  9.27E-9~ & 0 & ~~237  & 1.46E-12 & ~17838&~~1.91~& 62.2 \\

120729A  &0.80~~ &  2.08E-8~ & 0 & ~~106  & 3.06E-10 & ~~~864&~~2.65~& 6.98 \\
120907A  &0.97~~ &  9.07E-11 & 0 & 12691  & 2.38E-12 & ~64579&~~2.65~& 7.54 \\
141225A  &0.915~ &  3.04E-8~ & 0 & ~~542  & 1.63E-12 & ~23065&~~0.77~& 16.2 \\
160623A  &0.367~ &           & 0 &        & 2.40E-9~ & ~~6261&~~1.15~& 2.07 \\
160804A  &0.736~ &  2.81E-7~ & 0 & ~~136  & 3.47E-12 & ~77312&~~2.09~& 7.57 \\
170519A  &0.818~ &  3.06E-7  & 0 & ~~~52  & 3.05E-11 & ~11123&~~1.50~& 6.05 \\       
\hline
061110A  & 0.758~&  3.97E-9~ & 0 &~~~~33  & 2.33E-13 &158697~&~~1.26~& 19.7\\
         &       &  7.76E-11 & 0 & ~~~552 &          &       &       &      \\
080430~  & 0.768~&  2.56E-9~ & 0 & ~~~67 & 3.64E-13  &662118~&~~0.89~& 7.61\\       
         &       &  1.54E-11 & 0 & 264243 &          &        &      &       \\  
080916A  & 0.689~&  8.32E-8~ & 0 & ~~~~40 & 1.22E-12 &146113~&~~1.13~& 9.89\\
         &       &  2.19E-7~ & 0 & ~62767 &          &       &       &       \\
090814A  & 0.696~&  6.88E-8~ & 0 & ~~~125 & 8.46E-13 &~33274~& ~~0.90& 40.4\\
         &       &  2.61E-11 & 0 & ~~1332 &          &       &       &       \\
100621A  & 0.542~&  6.14E-7~ & 0 & ~~~~50 & 1.53E-11 &~66116~& ~~1.59& 5.31\\
         &       &  3.29E-10 & 0 & ~13750 &          &       &       &       \\
151027A  & 0.81~~&  2.64E-6  &91 & ~~~~19  & 5.37E-10&~~5563~& ~~1.84& 2.05  \\       
         &       &  2.37E-8  &173& ~~~~57  &         &       &       &         \\
\hline
\end{tabular}
\end{table*}

\begin{table*}
\caption{The parameters of the best fit
X-ray light curves of well-sampled  X-ray afterglow of
GRBs of known redshift  $z>4$ which could be well fit 
by Eq.(10) or by Eq.(4) plus Eq.(7).  Also
listed are the values of the period  of
the MSPs at birth, as estimated from the best fit values of
$F_{ps}$ and $t_b$. The index i is the pulse number}
\centering

\begin{tabular}{l l l l l l l l l l l}
\hline
\hline
~~SHB~~~&~~z~~~&~~~$F_i$~~~~&~$t_i$~&$\tau_i$~~&~~$F_{ps}$~~&~$t_b$&$\chi^2/dof$&~$P_0/\eta^{1/2}$~\\
        &     &$[erg/s\,cm^2\,]$&        &   ~[s]~  &$[erg/s\,cm^2]$~& ~[s]~& & [ms]\\
\hline

050505~ & 4.27~  &           &   &        & 3.56E-11 & ~14058~& 0.89 & 1.12 \\

050814~ & 5.77~  & 7.29E-12  &   & ~~140  & 2.99E-12 & ~41509~& 1.24  & 1.79 \\

050922B & 4.5~~  &           &   &        & 1.76E-12 & 180862~&       & 1.22 \\

060927~ & 5.47~  &  6.05E-11 &   & ~~138  & 2.45E-11 & ~~2403&~~0.78 & 3.06 \\

090205~ & 4.65   &  3.13E-11 & 0 & ~~206  & 4.46E-12 & ~10622&~~0.88 & 3.41 \\
   "     &        &  4.02E-12 & 0 & 34338  &$t_{2,0}=400s$&$\Delta_2=382s$&& \\

120521B & 6~~~~  &  4.45E-8  &   & ~~~41  & 9.39E-12 & ~~6064&~~1.08 & 2.56 \\

130606A & 5.913~ &           &   &        & 2.68E-11 & ~~8512&~~1.87 & 1.29 \\
  "     &        &  2.30E-7~ &   & ~~104  &$t_{1,0}=410s$&$\Delta_1=5484s$&& \\

140311A & 4.954~ &           &   &        & 1.38E-11 & ~17528&~~1.39 & 1.44 \\

140419A & 3.965~ &  3.25E-11 & 0 & ~8150  & 2.89E-11 & ~24340&~~1.26 & 1.00 \\
   "     &        &  7.62E-9~ & 0 & ~~~22  & $t_{2,0}=172s$& $\delta_2=51s$&&\\

140518A & 4.707~ &  8.27E-7~ & 0 & ~~~18  & 2.47E-11 & ~~3804&~~1.39 & 2.40 \\
    "   &        &  1.45E-10 &959& ~~~21  &          &       &       &      \\

140614A & 4.233~ &  5.79E-10 & 0 & ~~263  & 1.42E-11 & ~~7168&~~1.12 & 2.50 \\

151027B & 4.063  &  1.72E-7~ & 0 & ~~~48  & 8.63E-12 & ~27971&~~0.99 & 1.68 \\
\hline

\end{tabular}
\end{table*}

\end{document}